\documentclass[aps,prd,reprint,nofootinbib,twoside,notitlepage,superscriptaddress]{revtex4-1} 
\usepackage{graphicx}
\usepackage{euscript,amssymb}
\usepackage{amsfonts}
\usepackage{amssymb}
\usepackage{amsbsy}
\usepackage{amsmath}
\usepackage{nccmath}   
\usepackage{color}
\usepackage{bbold}
\usepackage{braket}
\usepackage{dsfont}
\usepackage[T3,T1]{fontenc}
\usepackage[section]{placeins}

\usepackage{subcaption} 
\usepackage{mathtools}

\mathtoolsset{showonlyrefs,showmanualtags}

\newcommand{\be}{\begin{equation}}
  \newcommand{\ee}{\end{equation}}
\newcommand{\ben}{\begin{eqnarray*}}
  \newcommand{\een}{\end{eqnarray*}}
\newcommand{\bea}{\begin{eqnarray}}
  \newcommand{\eea}{\end{eqnarray}}
\newcommand{\bdm}{\begin{displaymath}}
  \newcommand{\edm}{\end{displaymath}}
\newcommand{\ba}{\begin{align}}
  \newcommand{\ea}{\end{align}}

\newcommand{\sgn}{\text{sgn}\!}

\DeclareSymbolFont{tipa}{T3}{cmr}{m}{n}
\DeclareMathAccent{\invbreve}{\mathalpha}{tipa}{16}

\begin{document}

\title{Quantum Oppenheimer-Snyder model} 

\author{W{\l}odzimierz Piechocki} 
\email{wlodzimierz.piechocki@ncbj.gov.pl}
\affiliation{Department of Fundamental Research, National Centre for Nuclear
Research, Pasteura 7, 02-093 Warszawa, Poland}

\author{Tim Schmitz}
\email{tschmitz@thp.uni-koeln.de}
\affiliation{Institut f\"ur Theoretische Physik, Universit\"{a}t zu
K\"{o}ln, Z\"{u}lpicher Stra\ss e 77, 50937 K\"{o}ln, Germany}

\date{\today}

\begin{abstract}

We construct two reduced quantum theories for the Oppenheimer-Snyder model, respectively taking the point of view  of the comoving and the exterior stationary observer, using affine coherent states quantization. Investigations of the quantum corrected dynamics reveal that both observers can see a bounce, although for the exterior observer certain quantization ambiguities have to be chosen correctly. The minimal radius for this bounce as seen from the stationary observer is then shown to always be outside of the photon sphere. Possible avenues to lower this minimal radius and reclaim black holes as an intermediate state in the collapse are discussed. We demonstrate further that switching between the observers at the level of the quantum theories can be achieved by modifying the commutation relations.
	
\end{abstract}

\maketitle

\section{Introduction}

Singularities are an unavoidable feature of general relativity, but it is widely believed that they cannot occur in a full theory of quantum gravity. In the absence of a complete such theory we have to rely on specialized models to investigate how singularities could be avoided, what that means conceptually for the full theory, and where potential observational windows to it could be. In this paper we present such a model by quantizing the prototypical example for gravitational collapse to a black hole: the Oppenheimer-Snyder (OS) model for spherically symmetric, homogeneous dust collapse.

Quantization of the OS model has been discussed before with regard to its spectrum and possible singularity avoidance \cite{LundOS,CordaOS,PelegOSspectrum,PelegOS0,PelegOS}. Here we want to focus in particular on quantum corrected dynamics.

In Ref.\ \cite{MeOS} we have presented a canonical formulation of the OS model restricted to a flat interior, and in particular implemented the comoving and the exterior stationary observers explicitly. In this paper we want to quantize the OS model in this form, leading to two different deparametrized quantum theories, one for each observer. As we have already discussed in Ref.\ \cite{MeOS}, our investigation of a different collapse model in Ref.\ \cite{MeLTB} shows that quantum corrections as seen by the comoving observer will cause the dust cloud to bounce instead of collapsing to a singularity. Further we have found indications that from the point of view of the exterior observer the dust cloud can also bounce. In the following we want to investigate this in more detail.

These kind of quantum bounces avoiding classical singularities have in fact emerged in various approaches to quantum gravity. In addition to the results mentioned above, in the Wheeler-DeWitt approach null shells have been shown to bounce as well \cite{HajicekKieferNullShells,HajicekNullShells}. Similar results for null shells can also be obtained by using an effective one-loop action for quantum corrected gravity \cite{FrolovNullShell}. Loop quantum cosmology can yield bounces \cite{AshtekarCosmology} which have been discussed in the context of collapse models \cite{RovelliPlanckStars}, although the robustness of these bounces has been called into question recently \cite{BojowaldCritique}. Further there have been investigations of effective models for bouncing collapse based on the results above \cite{MalafarinaBounce,LiuBounce,AchourBounce1,AchourBounce2,AchourBounce3}.

There are many open questions regarding the consistency of bouncing collapse. The most important one is the lifetime of the temporary black hole produced by the collapse; in various approaches to bouncing collapse it has been shown that the transition from collapse to expansion has a lifetime proportional to the mass of the dust cloud \cite{AmbrusHajicekLifetime,ChristodoulouLifetime,ChristodoulouLifetime2,BarceloLifetime}. In general this would lead to extremely short lived black holes. There are proposals for mechanisms that could increase this lifetime \cite{ChristodoulouLifetime,BarceloBounce1} but no consensus has been reached so far. Related questions concern the behavior of the horizon during the bounce \cite{MalafarinaBounce, BambiBounce,  HajicekKieferNullShells, HajicekQuantumNullShells, BarceloBounce2, BarceloBounce3} and by what mechanism quantum gravitational effects even propagate to the horizon \cite{HaggardRovelliBounce, BarceloBounce3, BarceloBounce2}. For a review of bouncing collapse and all it entails see Ref.\ \cite{MalafarinaBounceRev}.

For most of these open questions, particularly for the lifetime, the point of view of the stationary observer is especially relevant. This is why we explicitly implement that observer here in the quantum theory; we hope to complement and build on our discussion of the comoving observer \cite{MeOS,MeLTB} and shed a bit more light on the aforementioned questions in bouncing collapse.

For some of the quantum theories producing bounces mentioned above the exterior observer is the preferred one, but this seems to be limited to the cases where null collapse was investigated. To our knowledge a quantization of a model for massive collapse from the point of view of the stationary observer has not been done before.

In addition, having access to two observers with their quantum theories will allow us to discuss the relationship between them. With this we can at the very least decide whether our method of implementing these observers by switching between them classically and quantizing the reduced theories is consistent. Further we can investigate how this switch could be realized within the quantum theories.

Due to the unusual form of the Hamiltonian for the stationary observer we apply a quantization method that might be unfamiliar to the reader: affine coherent states quantization (ACSQ). Coherent states quantization schemes rely on the identification of phase space with a Lie group. With the help of this Lie group one can construct a family of coherent states by letting a unitary irreducible representation of it on a Hilbert space act on some fixed state. Phase space functions are then mapped to operators on this Hilbert space by inserting them into the resolution of the identity that the coherent states bring with them. This allows also more complicated phase space functions to be paired up with operators, at least formally. Because the phase space here is a half plane we choose the affine group acting on the real line as the Lie group. For a more detailed explanation with some simple examples we refer the reader to \cite{BergeronCSQ,GazeauACSQ,AlmeidaACS}, or our introduction to the method below.

This quantization scheme has also been used in quantum cosmology, for example in Refs.\ \cite{BergeronBounce,BergeronMixmaster,GozdzBKL,GozdzBKL2}, and quite reliably seems to replace the singularity by a bounce.

Finally we want to mention that for simplicity we will restrict ourselves to the flat OS model, as we did in Ref.\ \cite{MeOS}. A generalization to curved Friedmann interiors would be interesting, but we do not believe that it will change our results significantly.

We proceed here as follows. In Sec.\ \ref{ch:chapter_2} we summarize the classical considerations from Ref.\ \cite{MeOS} that lead to the Hamiltonians we will discuss for the rest of the paper. In Sec.\ \ref{ch:chapter_3} we will quickly introduce the uninitiated reader to affine coherent states quantization, and apply this quantization scheme to the Hamiltonians relevant for the comoving observer in Sec.\ \ref{ch:chapter_4} and the stationary observer in Sec.\ \ref{ch:chapter_5}. We will primarily focus on the quantum corrected dynamics. How the two quantum theories can be related we discuss in Sec.\ \ref{ch:chapter_6}, before we finally conclude in Sec.\ \ref{ch:chapter_7}.

In the following we will use units where $G=c=1$.


\section{Canonical formulation of the OS model} \label{ch:chapter_2}

Here we will briefly recapitulate the canonical formulation of the OS model developed in Ref.\ \cite{MeOS}. We started from a general spherically symmetric spacetime. Following the usual Arnowitt-Deser-Misner procedure, the Einstein-Hilbert action for this class of spacetimes can be brought into canonical form, see  Ref.\ \cite{KucharSchwarzschild}. This involves arbitrarily foliating the spacetime by spatial hypersurfaces characterized by a fiducial coordinate frame with label time $t$ and radial coordinate $r$.

We included then into this canonical formulation Brown-Kucha\v{r} dust \cite{KucharBrownDust}, characterized by the canonical pair of dust proper time $\tau$ and dust density. To implement a discontinuity in the matter content we followed Ref.\ \cite{HajicekKijowskiFluid} and partially gauge fixed the fiducial coordinate frame such that the surface of the dust cloud we want to describe is always at a fixed $r=r_S$, splitting the spacetime into interior and exterior.

Let us first focus on the exterior, $r>r_S$, where the dust density vanishes. One can bring the constraints there into a fully deparametrizable form by promoting the Schwarzschild coordinates to phase space variables. This was done with the help of a series of canonical transformations developed in Ref.\ \cite{KucharSchwarzschild}. The constraints are then the momenta conjugate to the Schwarzschild coordinates. Both classically and in the quantum theory this simply tells us that the exterior is exactly Schwarzschild.

Through the fall-off behavior of the exterior canonical variables as they approach $r_S$ one can make sure that interior and exterior are smoothly matched. In particular we chose those conditions in such a way that the usual matching between a FLRW interior and a Schwarzschild exterior across $r=r_S$, see e.g.\ Ref.\ \cite{MeHamiltonian}, are appropriately reproduced in the canonical formalism.

Since we have fully deparametrized the exterior at the classical level, it is clear that the matching in this form will not hold up through quantization. In fact, it can not even be guaranteed that at the level of quantum corrections the matching can be made exact again by allowing a non-vanishing energy-momentum tensor on the collapsing body's surface, as was done e.g.\ in Ref.\ \cite{AchourBounce1}. We will briefly discuss at the end of this section how this restriction could be lifted in future work.

In the interior, $r<r_S$, where the dust density does not vanish, we imposed a further symmetry-reduction to homogeneity. To this end we restricted the dust density to be homogeneous, and the spacetime metric to be of FLRW form. This allowed us to integrate out the radial degree of freedom, giving us essentially a canonical formulation of Friedmann models with dust as matter, described by the scale factor $a$ and its momentum $p_a$, and the dust proper time $\tau$ and its momentum $P_\tau$. We further identified $P_\tau$ as the total mass of the collapsing body $M>0$.

The Hamiltonian constraint in the interior can then be expressed in the form $H+P_\tau$, and is hence deparametrizable with regard to dust proper time. $H$ therein is our first physical Hamiltonian. Restricting to flat interiors it is given by
\begin{equation}
H=-\frac{P^2}{2R} , \label{eq:comov_H}
\end{equation}
where $R = ar_S$ is the curvature radius of the dust cloud, and $P = p_a/r_S$ its canonical momentum. This Hamiltonian then describes the dust cloud from the point of view of the comoving observer.

The Hamiltonian is always negative, but this does not pose a problem. Classically it can be identified as $-M$, hence our Hamiltonian is negative but our notion of energy is not.

Note that the same Hamiltonian describes the Lema\^{i}tre-Tolman-Bondi model for inhomogeneous, spherically symmetric dust collapse, when one separates the dust cloud into a continuum of dust shells. This we have discussed in Ref.\ \cite{MeLTB}, where the above Hamiltonian was quantized using Dirac's canonical quantization. Large parts of this discussion carry over to the OS model, as already mentioned in Ref.\ \cite{MeOS}. For completeness we will also discuss this Hamiltonian using ACSQ and compare the results to Ref.\ \cite{MeLTB}. 

Additionally we promoted the Painlev\'{e}-Gullstrand coordinate transformation, which connects comoving time to Schwarzschild Killing time $T$, to a canonical transformation. Thereby we obtain $T$ as a canonical variable in the interior, given by
\begin{equation}
T =\tau  \pm 2\sqrt{2P_\tau}\left[ \sqrt{R} - \sqrt{\frac{P_\tau}{2}}\,\ln\left| \frac{\sqrt{R}+ \sqrt{2P_\tau}}{\sqrt{R}- \sqrt{2P_\tau}}\right|\right].
\end{equation}
Further we imposed $P_T=P_\tau$ to keep the convenient physical interpretation of this quantity as $M$. To make the canonical transformation complete one also needs to find a new momentum canonically conjugate to $R$ that has a vanishing Poisson bracket with $T$. This can be done by finding an appropriate generating function. The exact form of this new momentum is of no further importance here, details can be found in Ref.\ \cite{MeOS}. For simplicity we will also call this new momentum $P$.

The interior Hamiltonian constraint can then be brought into the form
\begin{equation}
H_T = P_T - \frac{R}{2}\begin{cases}
\tanh^2\frac{P}{R},~&R>2P_T\\
\coth^2\frac{P}{R},~&R<2P_T
\end{cases} . \label{eq:hamiltonian_interior2first}
\end{equation}
How the above can be deparametrized with regard to Killing time will be discussed in Sec.\ \ref{ch:chapter_5}. The resulting Hamiltonian will then describe the evolution of the dust cloud as seen from a stationary exterior observer.

It is straightforward to see that the constraint \eqref{eq:hamiltonian_interior2first} yields the expected behavior for the dust cloud. In particular we want to point out that as the momentum $P$ grows to infinity, the constraint goes to $R=2P_T=2M$: the dust cloud asymptotically approaches the horizon.

Note that the above cannot be easily adapted to the Lema\^{i}tre-Tolman-Bondi model. Switching the observer as above would apply to its outermost dust shell, but it is unclear how one would extend this to the other shells, since the Painlev\'{e}-Gullstrand coordinate transformation only applies up to the dust cloud's surface.

In summary we want to say that this setup to describe dust collapse canonically is convenient due its simplicity, but this simplicity comes with restrictions on the model. For the remainder of this section we want examine how one could go beyond some of these restrictions.

Firstly we want to discuss how our efforts could be adapted to non-flat interiors. Up to the Hamiltonian for the comoving observer such a generalization is straightforward, it simply adds a potential term linear in $R$ to Eq.\ \eqref{eq:comov_H}, with its sign determined by the interior's curvature.

It is when switching observers where the restriction to flat interiors simplifies matters considerably. The comoving time from the Painlev\'{e}-Gullstrand transformation describes comoving observers that start with zero velocity at infinity. Hence it can only be used for the flat case, where this initial condition is applicable. Dust clouds described by positively or negatively curved FLRW line elements fulfill instead other such conditions: the positive curvature case starts collapsing with zero velocity from a finite radius, and the negative curvature case has a non-zero velocity everywhere.

Coordinates that are applicable to these initial conditions are available, see Ref.\ \cite{MartelCoordinates} for the negative curvature case and Ref.\ \cite{GautreauCoordinates} for the positive curvature case. Especially the latter requires some additional care since it is only valid up to the initial radius of the collapsing body and would need to be extended beyond this region for our purposes. For simplicity we thus restricted ourselves to flat interiors.

Further restrictions are that the exterior is not allowed to differ from Schwarzschild, and that the interior is always homogeneous. In Ref.\ \cite{MeOS} we have already discussed indications that homogeneity could necessarily break near the classical singularity.  It seems to us that the most fruitful avenue for discussing these questions is to emulate the quantum corrected dynamics of the interior that we will derive here with an effective matter contribution. On the one hand, one can then investigate what exterior can be matched to it, as e.g.\ done in Refs.\ \cite{MalafarinaBounce,AchourBounce1,AchourBounce2, AchourBounce3}. On the other hand, one can implement inhomogeneity either perturbatively or by letting every dust shell in the cloud evolve separately, as discussed in Ref.\ \cite{MeLTB}. These considerations are outside of the scope of the present efforts.


\section{Affine coherent states quantization} \label{ch:chapter_3}

Coherent states can be defined with the help of a Lie group by letting a unitary irreducible representation of it on some Hilbert space act on an arbitrary state from that space \cite{PerelomovLieCS}. In coherent states quantization one can then use these coherent states to quantize a classical canonical theory by identifying the classical phase space with this Lie group. 

ACSQ is a coherent states quantization scheme for the Lie group being the affine group, and below we give a quick introduction to it. More details can be found e.g. in Refs.\ \cite{AlmeidaACS,GazeauACSQ,BergeronCSQ} and references therein.

We start from a canonical theory whose phase space is the half plane: One canonical variable $R$ is restricted to the positive half line, and its conjugate momentum $P$ takes values from all the reals. This phase space can be identified with the affine group of the real line, acting on $y\in\mathbb{R}$ as
\begin{equation}
(P,R)\cdot y=\frac{y}{R}+P .
\end{equation}
Note that this identification of the phase space with the affine group is highly ambiguous, and different choices lead to unitarily inequivalent quantum theories \cite{MeParameterizations}. We will make use of this ambiguity later, but for now we stick to the one given above. It has the advantage of behaving similarly to Dirac's prescription for canonical quantization, as we will see below. 

One can let this group act on the Hilbert space $\mathcal{H}=L^2(\mathbb{R}_+,dx)$ by using the representation
\begin{equation}
U(P,R)\cdot\psi(x)=\frac{e^{\frac{i}{\hbar}Px}}{\sqrt{R}}\psi\left( \tfrac{x}{R}\right) .
\end{equation}
Affine coherent states (ACS) can then be constructed as $\ket{P,R}=U(P,R)\ket{\Phi}$, where initially the so-called fiducial vector $\ket{\Phi}$ is a normalized element of the Hilbert space, fixed but arbitrary. Later more conditions on $\ket{\Phi}$ will be implemented as needed, such that certain numerical factors emerging from the quantum theory are finite. 

Choosing the fiducial vector can be seen as a quantization ambiguity in ACSQ, roughly corresponding to the factor ordering problem in Dirac's canonical quantization, as we will see later. In practice one usually considers a whole family of fiducial vectors to see how different choices impact the quantum theory.

With the help of the ACS one can now construct a resolution of the identity on the Hilbert space. To this end we need in addition to the ACS a measure on phase space invariant under the action of the affine group,
\begin{equation}
dR'dP'=dRdP,\quad(P',R')=(\pi,\rho)(P,R) .
\end{equation}
This measure is left invariant and allows us to write
\begin{equation}
\mathds{1}=\frac{1}{2\pi\hbar c^\Phi_{-1}}\int_{0}^\infty dR\int_{-\infty}^\infty dP~\ket{P,R}\bra{P,R} ,\label{eq:res_id}
\end{equation}
where we define
\begin{equation}
c^\Phi_\alpha=\int_{0}^{\infty}\frac{dx}{x^{2+\alpha}}~|\Phi(x)|^2 .
\end{equation}
Readers can convince themselves that this is indeed a resolution of the identity by confirming that it commutes with the irreducible representation $U(P,R)$. By Schur's lemma it then has to be at least a multiple of the identity. For a complete proof of \eqref{eq:res_id} see Ref.\ \cite{AlmeidaACS}.

In order for $c^\Phi_{-1}$ to be finite the first additional condition on $\Phi$ presents itself: $\Phi$ has to be square-integrable with regard to the measure $x^{-1}dx$. 

The quantization procedure now builds on \eqref{eq:res_id}. We assign functions $f(P,R)$ on phase space to operators acting on the Hilbert space by
\begin{equation}
f\mapsto \hat{f}=\frac{1}{2\pi\hbar c^\Phi_{-1}}\int_{0}^\infty dR\int_{-\infty}^\infty dP~f(P,R)~\ket{P,R}\bra{P,R} .
\end{equation}
The operators constructed in this way are automatically symmetric, and if the function $f$ is at least semi-bounded then $\hat{f}$ even has a self-adjoint extension in the Friedrich extension of the quadratic form defined as $\tilde{f}(\phi,\psi)=\braket{\phi|\hat{f}|\psi}$. For details see Ref.\ \cite{BergeronCSQ} and references therein.

As an example, consider how $\hat{R}$, $\hat{P}$ and the dilation $\hat{D}$, where $D=RP$, act:
\begin{align}
\hat{R}\psi(x)&=\frac{ c_{0}^\Phi}{ c_{-1}^\Phi}\,x\,\psi(x) ,\\
\hat{P}\psi(x)&=-i\hbar\psi(x)'-i\hbar\gamma\frac{\psi(x)}{x} ,\\
\hat{D}\psi(x)&=-i\hbar\frac{c^\Phi_0}{c^\Phi_{-1}}x\psi(x)'-i\hbar\lambda\psi(x) ,
\end{align}
where $\gamma$ and $\lambda$ are constants that depend on the fiducial vector. For details on the derivation see App.\ \ref{app:0}.
For real fiducial vectors $\gamma$ vanishes and $\lambda$ takes the value $c^\Phi_0/2c^\Phi_{-1}$, and the operators associated with position, momentum and dilation match those from Dirac quantization, aside from numerical factors. This direct correspondence between the two quantization schemes will not hold for more complicated phase space functions, as we will see later. 

Because the momentum operator matches its counterpart from Dirac quantization it cannot be made self-adjoint on the half line, and is thus not strictly speaking an observable. This is why we also explicitly consider the dilation here, which does not have that problem.

The canonical and affine commutation relations are also intact, regardless of $\Phi(x)$ being real,
\begin{align}
[\hat{R},\hat{P}]&=i\hbar\frac{ c_{0}^\Phi}{ c_{-1}^\Phi} ,\\
[\hat{R},\hat{D}]&=i\hbar\frac{ c_{0}^\Phi}{ c_{-1}^\Phi}\hat{R} ,
\end{align}
aside from the same numerical factor depending on $\ket{\Phi}$ as above. This factor can be made to disappear by a convenient choice of $\ket{\Phi}$, or simply absorbed by a reparametrization of $R$ in the affine group.

The last concept that we will need in the following is the notion of the lower symbol of a phase-space function. It is the expectation value of the operator associated with this function with regard to ACS,
\begin{align}
\check{f}&=\bra{P,R}\hat{f}\ket{P,R}\\&\hspace{-0.3em}=\frac{1}{2\pi \hbar c^\Phi_{-1}}\int_{0}^{\infty}dR'\int_{-\infty}^{\infty}dP'~f(P',R')~|\braket{P',R'|P,R}|^2 .
\end{align}
Consider once again as an example the lower symbols of position and momentum,
\begin{align}
\check{R}&=\frac{c_{-3}^\Phi c_{0}^\Phi}{ c_{-1}^\Phi}\, R ,\\
\check{P}&= P , \\
\check{D}&=D - i\hbar \alpha  ,\label{eq:lower_symbol_d}
\end{align}
where details can once again be found in App.\ \ref{app:0}. Aside from a prefactor and a constant imaginary shift $i\hbar\alpha$ in $\check{D}$, which vanishes for real fiducial vectors, these lower symbols then match the original phase space functions (apart from the rescaling of $R$ we already know from the commutation relations). As for the operators, this will generally not be the case for other functions, and the lower symbols will e.g.\ acquire additional terms as compared to their classical counterpart. 

In particular the lower symbol of the Hamiltonian has a special significance as it is often taken to generate quantum corrected time evolution in the classical phase space, see e.g.\ Refs.\ \cite{BergeronBounce,AlmeidaACS}. The viewpoint most commonly adopted seems to be that this quantum corrected sector is not to be identified within the full quantum theory, but is rather a theory of its own. This theory can be constructed by taking the action for full quantum mechanics and restricting the states that are to be varied to coherent states. Details can be found in Ref.\ \cite{KlauderEnhanced} and references therein. 

As we discuss in App.\ \ref{app:A}, one can also show that evolution of the coherent states according to the full Schr\"odinger equation implies evolution of their parameters $R$ and $P$ according to the lower symbol of the Hamiltonian as an approximation. The accuracy of the approximation thereby depends on the stability of the coherent states. In this way one can also obtain the quantum corrected phase space picture without leaving the full quantum theory. The difference between the two viewpoints is purely conceptual and does not have any operational consequences for the following.

Note that the same phase space picture emerges when one investigates coherent state propagators using a path integral approach, see e.g.\ Refs.\ \cite{WeissmanCS,KlauderCS} for a discussion of this for canonical coherent states. Conceptually this could be assigned to either of the viewpoints above.

Lastly we want to note that it is often useful to work with a centered fiducial vector, which means that it fulfills
\begin{align}
	1&=\braket{\Phi|\hat{x}|\Phi}=\int_0^\infty dx~x\,|\Phi(x)|^2=c^\Phi_{-3}\\
	0&=\braket{\Phi|\hat{d}_x|\Phi}=-i\hbar\int_0^\infty dx~\Phi^*(x) \left( x\frac{\partial}{\partial x}+\frac{1}{2} \right)\Phi(x)\\
\end{align}
The second condition can be realized by choosing a real fiducial vector. For more details see Ref.\ \cite{KlauderEnhanced}.


\section{Quantum theory for the comoving observer} \label{ch:chapter_4}

As recapitulated in Sec.\ \ref{ch:chapter_2}, the dynamics of the OS model from the point of view of the comoving observer can be described by the Hamiltonian
\begin{equation}
H=-\frac{P^2}{2R} .
\end{equation}
The Hamilton operator in the corresponding quantum theory accordingly takes the form
\begin{equation}
\hat{H}=\frac{1}{2\pi\hbar c_{-1}^\Phi}\int_{0}^{\infty}dR\int_{-\infty}^{\infty}dP~H(P,R)~\ket{P,R}\bra{P,R} ,\label{eq:hamiltonian0}
\end{equation}
and acts on wave functions as
\begin{align}
\hat{H}\psi(x)&=-\frac{1}{2\pi\hbar c_{-1}^\Phi}\int_{0}^{\infty}dR\int_{-\infty}^{\infty}dP\int_{0}^{\infty}dy~\frac{P^2}{2R^2} \nonumber\\&\quad\quad e^{\frac{i}{\hbar}(x-y)P}\Phi\!\left(\tfrac{x}{R}\right)\Phi^*\!\left(\tfrac{y}{R}\right)~\psi(y)\\
&=\frac{\hbar^2}{2 c_{-1}^\Phi}\int_{0}^{\infty}\frac{dR}{R^2}~\Phi\!\left(\tfrac{x}{R}\right)~\frac{\partial^2}{\partial x^2}~ \Phi^*\!\left(\tfrac{x}{R}\right)\psi(x),
\end{align}
where we have used the identity $\int_{-\infty}^{\infty}dP~P^2 e^{i(x-y)P}=-2\pi\,\delta''(x-y)$. 

For simplicity we will only consider real fiducial vectors. The Hamiltonian can then be simplified by partial integration. We understand our assumption that no boundary terms emerge from these partial integrations as additional boundary conditions on the fiducial vector. Finally we arrive at
\begin{align}
\hat{H}\psi(x)&=\frac{\hbar^2}{2 c_{-1}^\Phi}\left(\frac{1-c^{\Phi'}_{-4}}{x^3}\,\psi(x) -\frac{1}{x^2}\,\psi'(x) + \frac{1}{x}\,\psi''(x)\right) \\
&=\frac{\hbar^2}{2 c_{-1}^\Phi}x^{-1\mp\sqrt{c^{\Phi'}_{-4}}}\frac{\partial}{\partial x}x^{1\pm 2\sqrt{c^{\Phi'}_{-4}}}\frac{\partial}{\partial x}x^{-1\mp\sqrt{c^{\Phi'}_{-4}}}\,\psi(x) .
\end{align}
It is obvious that the Hamiltonian is the same as the one obtained from Dirac's quantization in Ref.\ \cite{MeLTB}, apart from an irrelevant prefactor, restricted to a specific class of factor orderings ($a+2b=1$ in the notation of Ref.\ \cite{MeLTB}). There a Hilbert space with measure $x^{1-a-2b}dx$ was investigated, which then matches under the aforementioned constraint our Hilbert space here. Therefore we suspect that one can achieve a perfect match between the two approaches by allowing measures in our Hilbert space different from the standard one, but this is not necessary for our purpose.

Self-adjoint extensions of the Hamiltonian were discussed in Ref.\ \cite{MeLTB}, as well as solutions to the corresponding Schr\"odinger equation. It was further shown that the norm squared of those solutions behaves towards $x=0$ to leading order as $x^{2+2\sqrt{c^{\Phi'}_{-4}}}$, as long as $4\, c^{\Phi'}_{-4}\geq9$ or one chooses a specific self-adjoint extension of the Hamiltonian ($\theta=\pi$ in the notation of Ref.\ \cite{MeLTB}). Should these conditions not be fulfilled, the solutions behave instead like $x^{2-2\sqrt{c^{\Phi'}_{-4}}}$. Note that the cases where $\frac{2}{3}\sqrt{c^{\Phi'}_{-4}}\in\mathbb{Z}$ had to be excluded from the discussion. Interpreting the norm squared of wave functions as the probability distribution for $R$ in analogy to Dirac's canonical quantization, the above shows that the classical singularity is avoided in the quantum theory when one either chooses the $\theta=\pi$ self-adjoint extension, or the fiducial vector such that $4\,c^{\Phi'}_{-4}\geq9$ or $c^{\Phi'}_{-4}<1$.

We have further shown in Ref.\ \cite{MeLTB} that the spectrum of the Hamiltonian above is the negative real line, potentially with an additional positive eigenvalue. This positive eigenvalue only occurs for some self-adjoint extensions, and its value strongly depends on this extension and the chosen factor ordering. We can hence exclude it as unphysical. Since the Hamiltonian is interpreted as the negative mass of the dust cloud, as discussed in Sec.\ \ref{ch:chapter_2}, the energy of the system is therefore positive definite.

In Ref.\ \cite{MeLTB} we further discussed quantum corrected dynamics of the system by considering expectation values of a wave packet. Here we want to compare these results to the quantum corrected dynamics obtained from the lower symbol of the Hamiltonian.

Let us find this lower symbol, keeping in mind we have restricted ourselves to real fiducial vectors:
\begin{align}
\check{H}&=\bra{P,R}\hat{H}\ket{P,R}\\
&=\frac{\hbar^2}{2 R c_{-1}^\Phi}\int_{0}^{\infty}dx~ e^{-\frac{i}{\hbar}Px} \Phi\!\left( \tfrac{x}{R}\right)\\&\quad\quad   \left(\frac{1-c^{\Phi'}_{-4}}{x^3} -\frac{1}{x^2}\frac{\partial}{\partial x} + \frac{1}{x}\frac{\partial^2}{\partial x^2}\right) e^{\frac{i}{\hbar}Px} \Phi\!\left( \tfrac{x}{R}\right) \\
&=-\frac{P^2}{2R}-\frac{c^{\Phi'}_{-1}+c^{\Phi}_{1}(c^{\Phi'}_{-4}-1)}{2c^{\Phi}_{-1}}\frac{\hbar^2}{R^3} . \label{eq:eff_hamiltonian}
\end{align}
$\check{H}$ consists of the classical Hamiltonian and an additional potential, vanishing in the classical limit $\hbar\to0$. Note that for more complicated phase space functions the classical limit is obtained by also taking a limit in the chosen family of fiducial vectors, as we will see in Sec.\ \ref{ch:chapter_5}.

The potential in \eqref{eq:eff_hamiltonian}  will always be repulsive. $c^{\Phi'}_{-4}$ is positive and we can rewrite
\begin{equation}
c^{\Phi'}_{-1}-c^\Phi_1=\int_0^\infty \frac{dx}{x^3}~\left(\Phi(x)-x\Phi(x)' \right)^2\geq 0. 
\end{equation}
In general, this identity holds up to a boundary term due to partial integration. However, since we have already implicitly restricted the behavior of the fiducial vector for $x\to0$ and $x\to\infty$ by assuming the finiteness of the $c^\Phi_\alpha$'s appearing above, this boundary term vanishes. It follows that the numerical factor in front of the effective potential in \eqref{eq:eff_hamiltonian} is always positive.

In summary we can say that ACS evolve as determined by an effective Hamiltonian
\begin{equation}
\check{H}=-\frac{P^2}{2R}-\frac{\hbar^2\delta}{R^3} ,\label{eq:eff_hamiltonian3}
\end{equation}
where the factor $\delta>0$ depends on the choice of fiducial vector. Note that this Hamiltonian matches the one for Friedmann models found in Ref.\ \cite{KlauderEnhanced} using Klauder's enhanced quantization. Further, in Ref. \cite{CasadioOSBounce} a quantum corrected Hamilton-Jacobi equation equivalent to \eqref{eq:eff_hamiltonian3} was found from a Born-Oppenheimer approximation on minisuperspace, albeit with a slightly different repulsive potential.

To solve the equations of motion given by \eqref{eq:eff_hamiltonian3}, we take into account that the Hamiltonian itself is a constant of motion, $-M$ with $M>0$, and find
\begin{equation}
\left( \frac{d R}{d \tau}\right)^2 = \frac{2M}{R} - \frac{2\hbar^2\delta}{R^4} ,
\end{equation}
giving the solution
\begin{equation}
R(\tau)=\left(\frac{\hbar^2\delta}{M} + \frac{9M}{2}(\tau-\tau_0)^2 \right)^\frac{1}{3} .
\end{equation}
Remarkably, this exactly matches the quantum corrected dust trajectories in Ref.\ \cite{MeLTB}: For large $R$, the trajectory reproduces the classical expanding/collapsing trajectories, which are connected by a bounce replacing the collapse to a singularity. The minimal radius depends on quantization ambiguities, here the fiducial vector, and scales with the energy as $M^{-\frac{1}{3}}$.

Depending on the factor $\delta$, the dust cloud can temporarily fall behind its horizon $R=2M$ and a stage of the collapse similar to a black hole is reached. As already mentioned in the introduction, it is important to know how long this stage lasts. We can estimate this lifetime by
\begin{equation}
	\Delta\tau=\tau_+ - \tau_- ,
\end{equation}
where $R(\tau_\pm)=2M$. This leads to the expression
\begin{equation}
	\Delta\tau=\frac{8M}{3}\sqrt{1-\left(\frac{R_0}{2M} \right)^3 },
\end{equation}
where $R_0=R(\tau_0)$. When the minimal radius $R_0$ is significantly smaller than the horizon we recover the result from Ref.\ \cite{MeLTB}: from the point of view of the comoving observer, the lifetime of the black hole stage is proportional to the mass of the dust cloud.

Also concerning the quantum corrected trajectory in momentum space $P(\tau)$ we have the same agreement between the two approaches. As follows from the Hamiltonian equations of motion for \eqref{eq:eff_hamiltonian3},
\begin{equation}
P=-R\,\frac{d R}{d \tau}=-\frac{3M}{R}(\tau-\tau_0) .
\end{equation}

In conclusion, we can say that the ACSQ of this specific Hamiltonian quite successfully reproduces the results of the usual Dirac quantization discussed in Ref.\ \cite{MeLTB}. Finding the quantum corrected dynamics of the system is even easier in ACSQ, since one only has to compute the lower symbol of the Hamiltonian. 


\section{Quantum theory for the stationary observer} \label{ch:chapter_5}

From the point of view of the stationary observer we have the Hamiltonian constraint
\begin{equation}
H_T = P_T - \frac{R}{2}\begin{cases}
\tanh^2\frac{P}{R},~&R>2P_T\\
\coth^2\frac{P}{R},~&R<2P_T
\end{cases} .\label{eq:hamiltonian_interior2}
\end{equation}
We first want to discuss how we can arrive at a deparametrized quantum theory with regard to $T$. In the current form \eqref{eq:hamiltonian_interior2}, the effective Hamiltonian still depends on $P_T$ through the position of the split between inside and outside of the horizon. As already noted in Ref.\ \cite{MeOS}, having the explicit split in the constraint is unnecessary, since the conditions $R<2P_T$ and $R>2P_T$ are implemented automatically by the constraint itself: On the constraint surface we have $2P_T= R \tanh^2\frac{P}{R}<R$ and respectively on the inside. We can then identify an effective Hamiltonian, albeit a multivalued one,
\begin{equation}
H = - \frac{R}{2}\begin{cases}
\tanh^2\frac{P}{R}\\
\coth^2\frac{P}{R}
\end{cases} . \label{eq:hamiltonian_Killing_time}
\end{equation}

It is furthermore convenient to bring the Hamiltonian into a slightly different form with a simple canonical transformation: we introduce $A=\frac{1}{2}R^2$, proportional to the surface area of the dust cloud, and its canonical momentum $P_A=P/R$. The Hamiltonian then takes the form
\begin{equation}
H = - \sqrt{\frac{A}{2}}\begin{cases}
\tanh^2P_A\\
\coth^2P_A
\end{cases} .\label{eq:2effectiveH}
\end{equation}

If we want to directly quantize this Hamiltonian, we also have to allow multivalued quantum states, both branches evolving with respect to a branch of the Hamiltonian operator. Such a construction has been discussed before in Ref.\ \cite{ShapereMultivalued}. There it has been shown that in such a multivalued quantum theory unitary time evolution can still be implemented with boundary conditions on the wave function at the Hamiltonian's branching points. Because we are mainly interested in the quantum corrected dynamics of the theory, we will not explicitly implement the construction from Ref.\ \cite{ShapereMultivalued}, but it would certainly be possible. The branching points for our Hamiltonian would be at $P_A\to\pm\infty$. Alternatively one can view the following as a quantization of two completely different systems. As it turns out, what specific viewpoint is taken does not make much of a difference.

There are other approaches to quantizing multivalued Hamiltonians: Adding constraints implementing the velocities as phase space coordinates, eventually leading to the use of Dirac brackets, has been used e.g.\ in Refs.\ \cite{LiuMultivalued,RuzMultivalued}. Alternatively, in Ref.\ \cite{HenneauxMultivalued} an effective Hamiltonian was found as a sum of the different branches of the original multivalued one by considering the path integral. Unfortunately all these methods rely on the multivalued Hamiltonian emerging from a single-valued Lagrangian, as e.g.\ in modified gravity where the velocities occur in powers higher than two. This is not the case for our Hamiltonian, leaving us only with the approach from Ref.\ \cite{ShapereMultivalued}.

As is apparent, even with the problem of deparametrization out of the way this Hamiltonian still presents some challenges if one wants to follow Dirac's prescription for canonical quantization. We will instead employ ACSQ to tackle this problem. Earlier we have seen that there at least formally every phase space function has an associated operator, so the fact that the hyperbolic tangent and cotangent both do not have a Taylor series that is defined everywhere is not a problem. Furthermore, since both branches of the Hamiltonian are semi-bounded the resulting operators will have a self-adjoint extension. The fact that the Hamiltonian has a complicated dependency on the momentum is hence not a problem, at least on the formal level.

Unfortunately the Hamilton operators as acting in position space are the sum of a multiplicative and an unbounded integral operator. It is thus quite challenging to find eigenfunctions or make statements about the spectrum, but we can still discuss the quantum corrected dynamics of the system. To this end we will consider the lower symbols of both branches of the Hamiltonian separately from each other.

We will call the outside branch of the Hamiltonian $H_+(P_A,A)$ and the inside branch $H_-(P_A,A)$. Their respective lower symbols are
\begin{align}
\check{H}_+&=-\frac{1}{2\pi\hbar c^\Phi_{-1}}\int_{0}^\infty d\bar{A}\int_{-\infty}^\infty d\bar{P}_A~\left|\braket{\bar{P}_A,\bar{A}|P_A,A}\right|^2\\ &\quad\quad  \left[ \sqrt{\frac{\bar{A}}{2}} \left( \tanh^2 \bar{P}_A-1\right) +   \sqrt{\frac{\bar{A}}{2}} \right]
\end{align}
and
\begin{align}
\check{H}_-&=-\frac{1}{2\pi\hbar c^\Phi_{-1}}\int_{0}^\infty d\bar{A}\int_{-\infty}^\infty d\bar{P}_A~\left|\braket{\bar{P}_A,\bar{A}|P_A,A}\right|^2\\ &\quad\quad \left[ \sqrt{\frac{\bar{A}}{2}} \left( \coth^2 \bar{P}_A -1 \right) + \sqrt{\frac{\bar{A}}{2}} \right] ,
\end{align}
where
\begin{multline}
\left|\braket{\bar{P}_A,\bar{A}|P_A,A}\right|^2=\int_{0}^{\infty}dx\int_{0}^{\infty}dy~\frac{ e^{ \frac{i}{\hbar}\left(P_A - \bar{P}_A\right) \left(  x - y\right) }}{A\,\bar{A}}\\ \Phi\left( \tfrac{x}{A}\right)\Phi^*\left( \tfrac{x}{\bar{A}}\right)\Phi^*\left( \tfrac{y}{A}\right)\Phi\left( \tfrac{y}{\bar{A}}\right).
\end{multline}
We have rewritten the Hamiltonian slightly to make use of the following identities:
\begin{align}
\int_{-\infty}^\infty &d\bar{P}_A~ \left( \tanh^2\bar{P}_A- 1\right) e^{- \frac{i}{\hbar}\bar{P}_A\left(  x - y\right) }\\&=-\int_{-\infty}^\infty d\bar{P}_A~ \frac{e^{- \frac{i}{\hbar}\bar{P}_A\left(  x - y\right) }}{\cosh^2\bar{P}_A}=-\frac{\frac{\pi }{\hbar}\left(  x - y\right)}{\sinh\left(\frac{\pi }{2\hbar}\left(  x - y\right) \right) } ,\label{eq:2useful_integral1}\\
\int_{-\infty}^\infty &d\bar{P}_A~ \left( \coth^2\bar{P}_A- 1\right)  e^{ -\frac{i}{\hbar}\bar{P}_A\left(  x - y\right) }\\&=\int_{-\infty}^\infty d\bar{P}_A~ \frac{e^{- \frac{i}{\hbar}\bar{P}_A\left(  x - y\right) }}{\sinh^2\bar{P}_A} =- \frac{\frac{\pi }{\hbar}\left(  x - y\right)}{\tanh\left(\frac{\pi }{2\hbar}\left(  x - y\right) \right) }\label{eq:2useful_integral2}  .
\end{align}
The first Fourier transformation can straightforwardly be obtained by contour integration. In computing the second integral one has to be a bit more careful, because the integrand diverges for $\bar{P}_A\to0$. We have regularized it by performing two contour integrations, one over a contour including this divergence and one excluding it, and averaging over the results. The full derivation can be found in App.\ \ref{app:B}.

The lower symbols then take the form
\begin{multline}
\check{H}_\pm=-\frac{c_{-\frac{5}{2}}^\Phi c_{-\frac{1}{2}}^\Phi}{ c_{-1}^\Phi} \sqrt{\frac{A}{2}} +\frac{1}{2\sqrt{2}\hbar^2 c^\Phi_{-1} \,A}\\\int_{0}^\infty \frac{d\bar{A}}{\sqrt{\bar{A}}}\int_{0}^\infty dx\int_{0}^\infty dy~  \frac{ x - y }{ F_\pm\left(\frac{\pi }{2\hbar}\left(  x - y\right) \right) } \\ e^{\frac{i}{\hbar}P_A\left(x-y \right) } \Phi\left( \tfrac{x}{A}\right) \Phi^*\left( \tfrac{y}{A}\right) \Phi^*\left( \tfrac{x}{\bar{A}}\right) \Phi\left( \tfrac{y}{\bar{A}}\right) \label{eq:hamiltonian_generic} ,
\end{multline}
where
\begin{equation}
F_+(x)=\sinh(x)\quad\text{and}\quad F_-(x)=\tanh(x)
\end{equation}

To progress any further we need to specify a fiducial vector. A convenient choice, borrowed from Ref.\  \cite{KlauderEnhanced}, is 
\begin{equation}
\Phi(x)=\frac{(2\beta)^\beta}{\sqrt{\Gamma(2\beta)}} x^{\beta-\frac{1}{2}} e^{-\beta x} ,
\end{equation}
where $\beta$ is a positive real parameter. The relevant constants are then
\begin{align}
c^\Phi_{-1}&=\frac{2\beta}{2\beta-1} ,\\
c^\Phi_{-\frac{1}{2}}&=2\sqrt{2}\beta^\frac{3}{2}\frac{\Gamma\!\left(2\beta -\frac{3}{2} \right) }{\Gamma\!\left(2\beta \right)} ,\\
c^\Phi_{-\frac{5}{2}}&=\frac{\Gamma\!\left(2\beta +\frac{1}{2} \right) }{\sqrt{2\beta}\Gamma\!\left(2\beta \right)} ,\\
c^\Phi_{-3}&=1 .
\end{align}
For these constants to be finite we have to impose $\beta>\frac{3}{4}$. Note that this fiducial vector is centered.

We can then perform the $\bar{A}$ integration,
\begin{align}
\int_0^\infty\frac{d\bar{A}}{\sqrt{\bar{A}}} ~& \Phi^*\left( \tfrac{x}{\bar{A}}\right) \Phi\left( \tfrac{y}{\bar{A}}\right)\\ &= \frac{(2\beta)^{2\beta}}{\Gamma(2\beta)} \left( x y \right)^{\beta-\frac{1}{2}} \int_0^\infty\frac{d\bar{A}}{\bar{A}^{2\beta-\frac{1}{2}}}~ e^{-\frac{\beta}{\bar{A}} \left( x + y\right) }\nonumber \\
&=\frac{2^{2\beta} \beta^\frac{3}{2}}{\Gamma(2\beta)} \frac{\left( x y \right)^{\beta-\frac{1}{2}}}{\left( x+y\right)^{2\beta-\frac{3}{2}}  } \Gamma\!\left(2\beta -\tfrac{3}{2} \right)  .
\end{align}
This gives as the lower symbols
\begin{multline}
\frac{ \Gamma\!\left(2\beta -1 \right) \Gamma\!\left(2\beta\right)}{\Gamma\!\left(2\beta +\tfrac{1}{2} \right) \Gamma\!\left(2\beta -\tfrac{3}{2} \right)}~\check{H}_\pm(P_A,A)\\=- \sqrt{\frac{A}{2}} +\frac{2^{4\beta-\frac{5}{2}} \beta^{2\beta+\frac{1}{2}} }{\hbar^2 \Gamma\!\left(2\beta +\tfrac{1}{2} \right) \,A^{2\beta}}\int_{0}^\infty dx\int_{0}^\infty dy~ \\ \frac{ x - y }{ F_\pm\left(\frac{\pi }{2\hbar}\left(  x - y\right) \right) } \frac{\left( x y \right)^{2\beta-1}}{\left( x+y\right)^{2\beta-\frac{3}{2}}  } e^{\frac{i}{\hbar}P_A\left(x-y \right) -\frac{\beta}{A}\left( x+y\right)  }    \label{eq:hamiltonian_outinside} .
\end{multline}

\begin{figure}
	\centering
	\begin{subfigure}{0.45\textwidth}
		\centering
		\includegraphics[width=\textwidth]{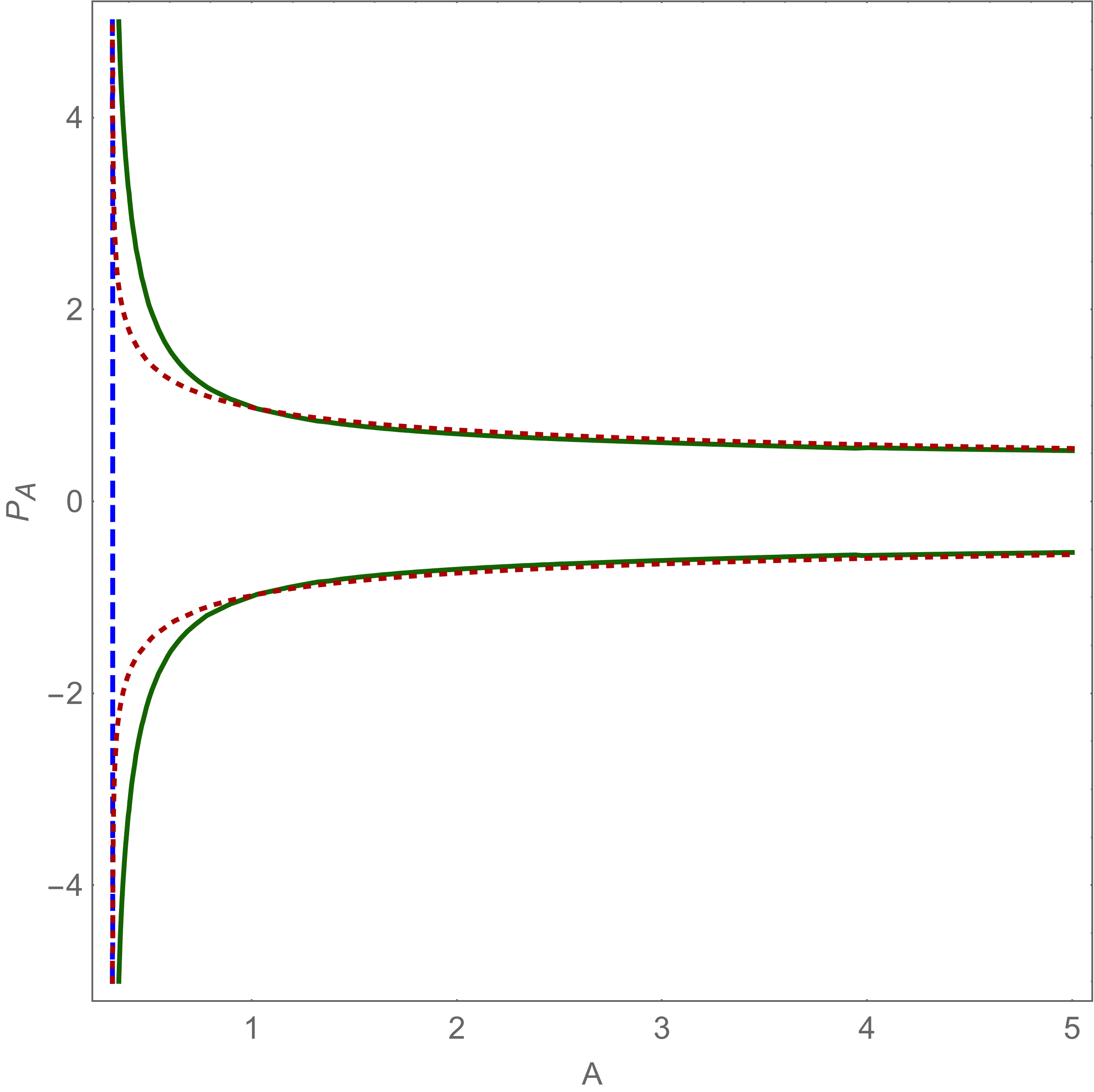}
		\caption{$M=0.4$ and $\beta=1$.}
	\end{subfigure}
	\begin{subfigure}{0.45\textwidth}
		\centering
		\includegraphics[width=\textwidth]{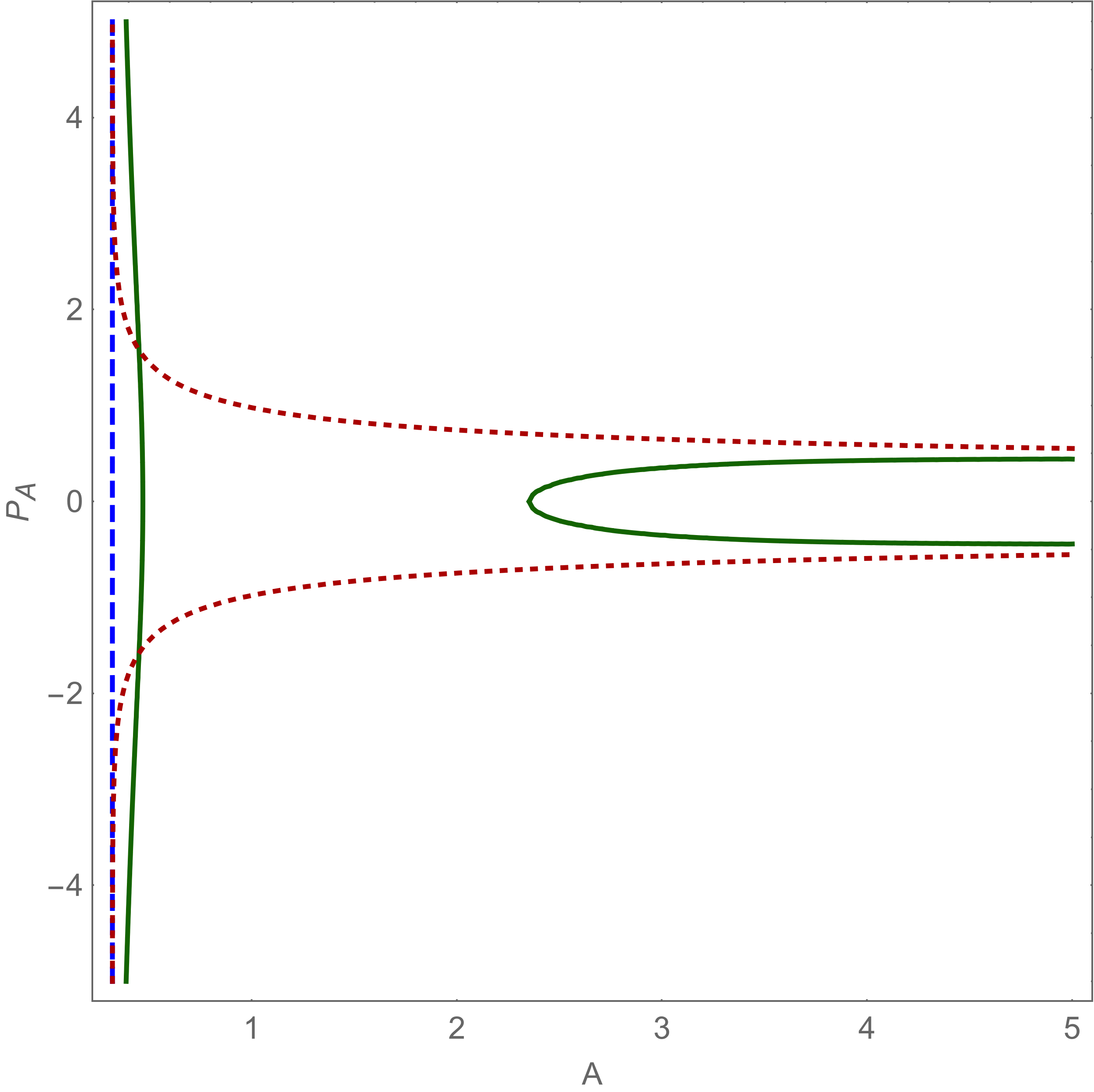}
		\caption{$M=0.4$ and $\beta=5$.}
	\end{subfigure}
	\caption{The quantum corrected phase space portraits for the outside branch of the Hamiltonian as given by \eqref{eq:phase_space_portrait} (full green line) as compared to their classical counterpart (dotted red line), given by $M=\sqrt{\frac{A}{2}}\tanh^2 P_A$, and the horizon $M=\sqrt{\frac{A}{2}}$ (dashed blue line), in Planck units.}
	\label{fig:phase_space_portraits_outside}
\end{figure}

\begin{figure*}
	\begin{subfigure}{0.45\textwidth}
		\centering
		\includegraphics[width=\textwidth]{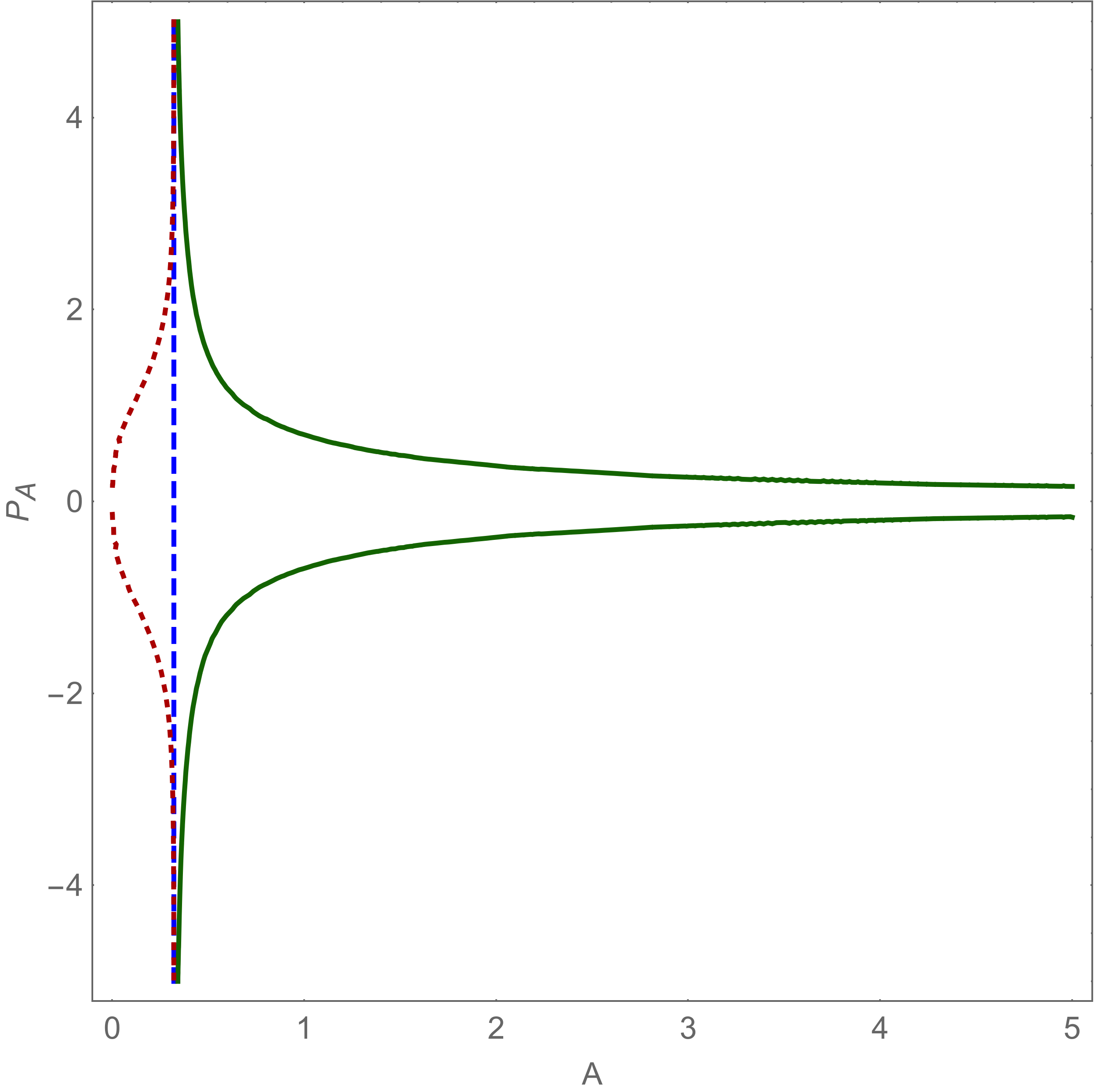}
		\caption{$M=0.4$ and $\beta=1$.}
	\end{subfigure}
	\vspace{1em}
	\begin{subfigure}{0.45\textwidth}
		\centering
		\includegraphics[width=\textwidth]{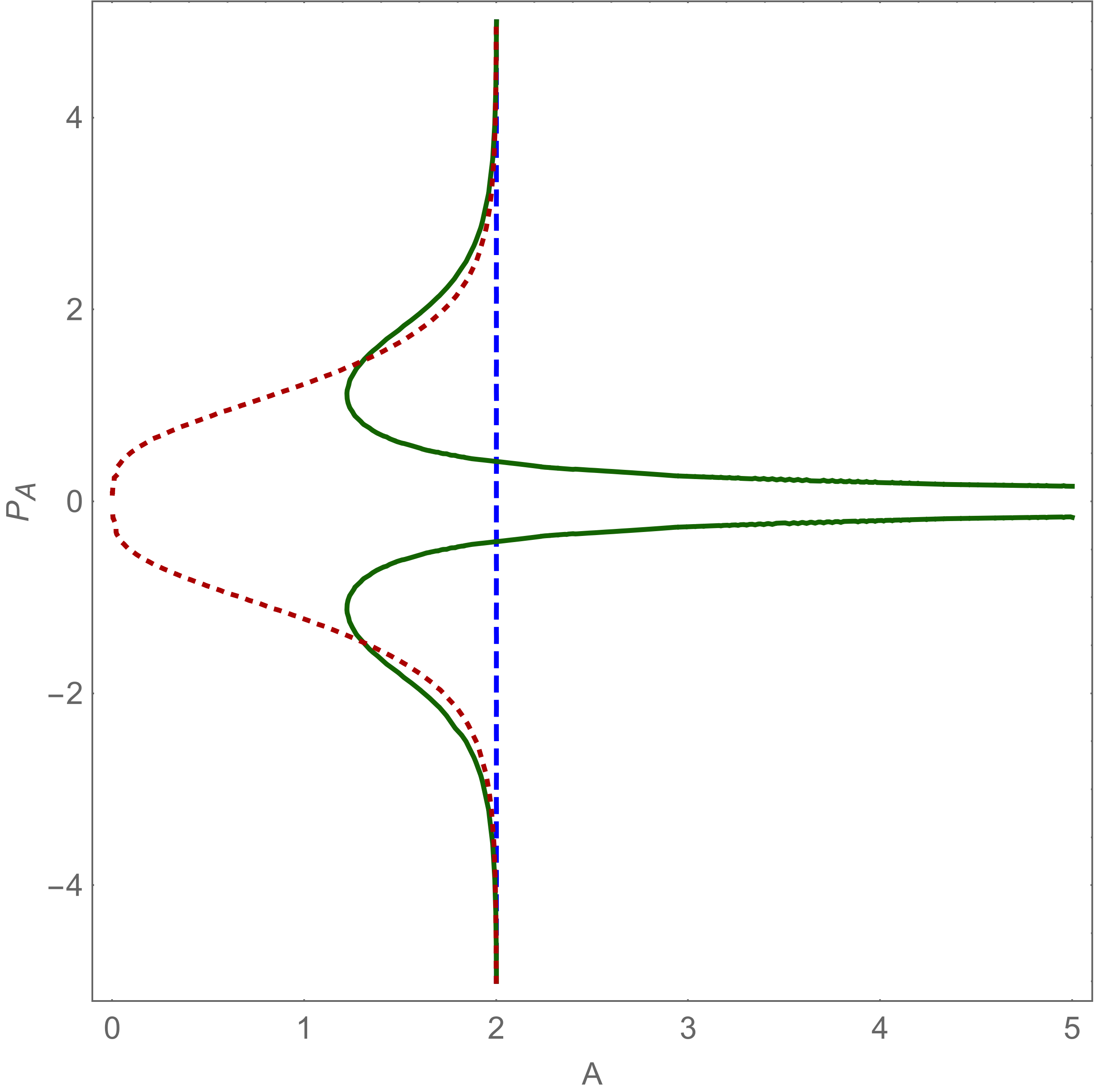}
		\caption{$M=1$ and $\beta=1$.}
	\end{subfigure}
	\begin{subfigure}{0.45\textwidth}
		\centering
		\includegraphics[width=\textwidth]{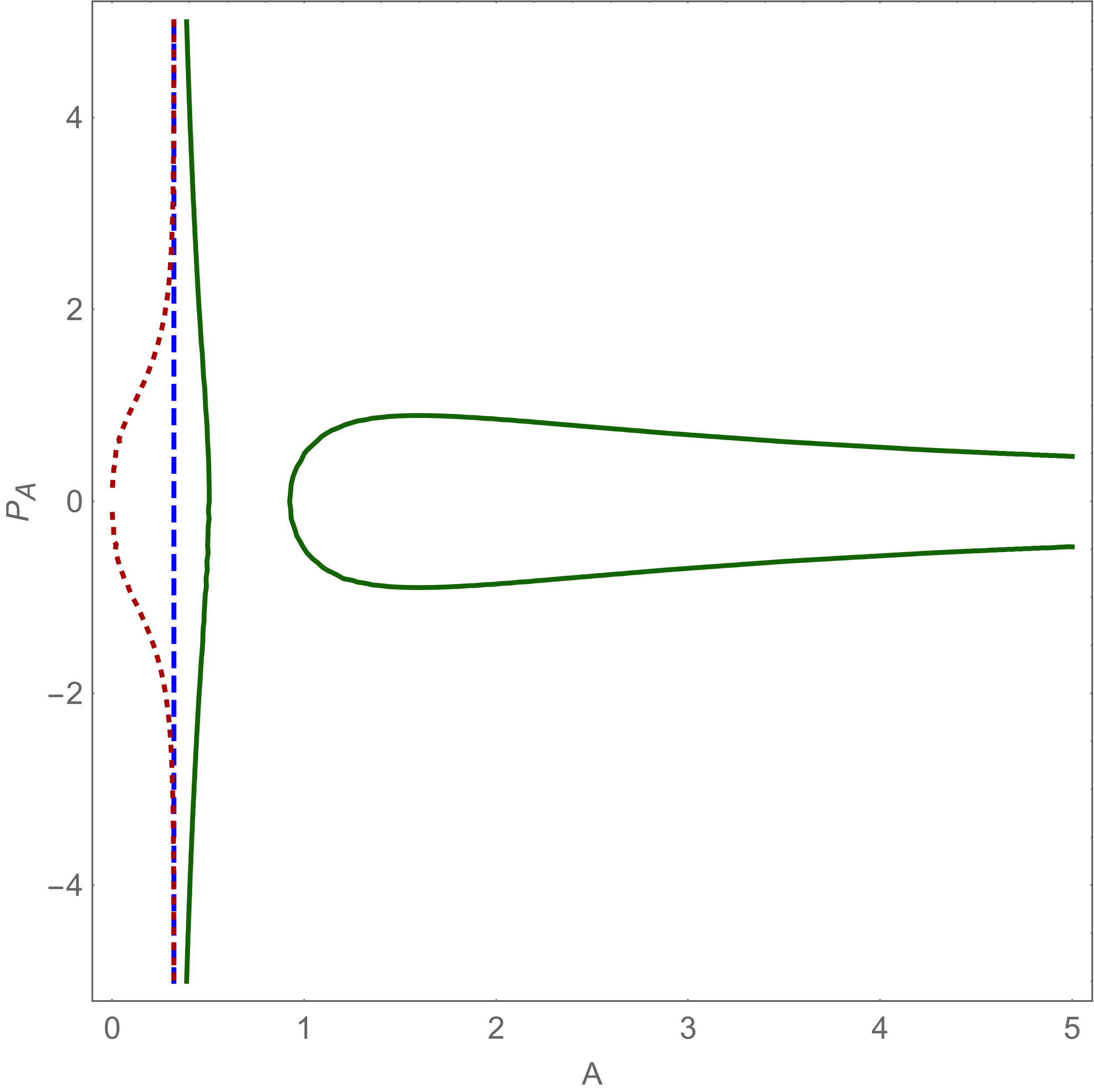}
		\caption{$M=0.4$ and $\beta=5$.}
	\end{subfigure}
	\caption{The quantum corrected phase space portraits for the inside branch of the Hamiltonian as given by \eqref{eq:phase_space_portrait} (full green line) as compared to their classical counterpart (dotted red line), given by $M=\sqrt{\frac{A}{2}}\coth^2 P_A$, and the horizon $M=\sqrt{\frac{A}{2}}$ (dashed blue line), in Planck units.\\\vspace{5em}}
	\label{fig:phase_space_portraits_inside}
\end{figure*}

These quantum corrected Hamiltonians can be investigated numerically. Before solving the equations of motion we can first consider phase space portraits. To this end we identify as the mass of the quantum corrected dust cloud
\begin{equation}
M=-\frac{ \Gamma\!\left(2\beta -1 \right) \Gamma\!\left(2\beta\right)}{\Gamma\!\left(2\beta +\tfrac{1}{2} \right) \Gamma\!\left(2\beta -\tfrac{3}{2} \right)}~\check{H}_\pm(P_A,A) .\label{eq:phase_space_portrait}
\end{equation}
After all, for $|P_A|\to\infty$ the second term in the Hamiltonians vanishes, since Fourier transforms of integrable functions vanish at infinity, leading to $A=2M^2$; the classical horizon is still present at least kinematically, so we can find the mass of the dust cloud from the area of this horizon. See Fig.\ \ref{fig:phase_space_portraits_outside} and Fig.\ \ref{fig:phase_space_portraits_inside} for the phase space portraits. 

Firstly we want to note that the figures we show here are restricted to masses close to the Planck scale, and similarly for the parameter $\beta$, since this makes the numerical computations considerably more stable. We will show analytically at the end of this section that the results discussed in the following also apply to astrophysical scales, at least when one chooses $\beta$ accordingly.

\begin{figure*}
	\centering
	\begin{subfigure}{0.45\textwidth}
		\centering
		\includegraphics[width=\textwidth]{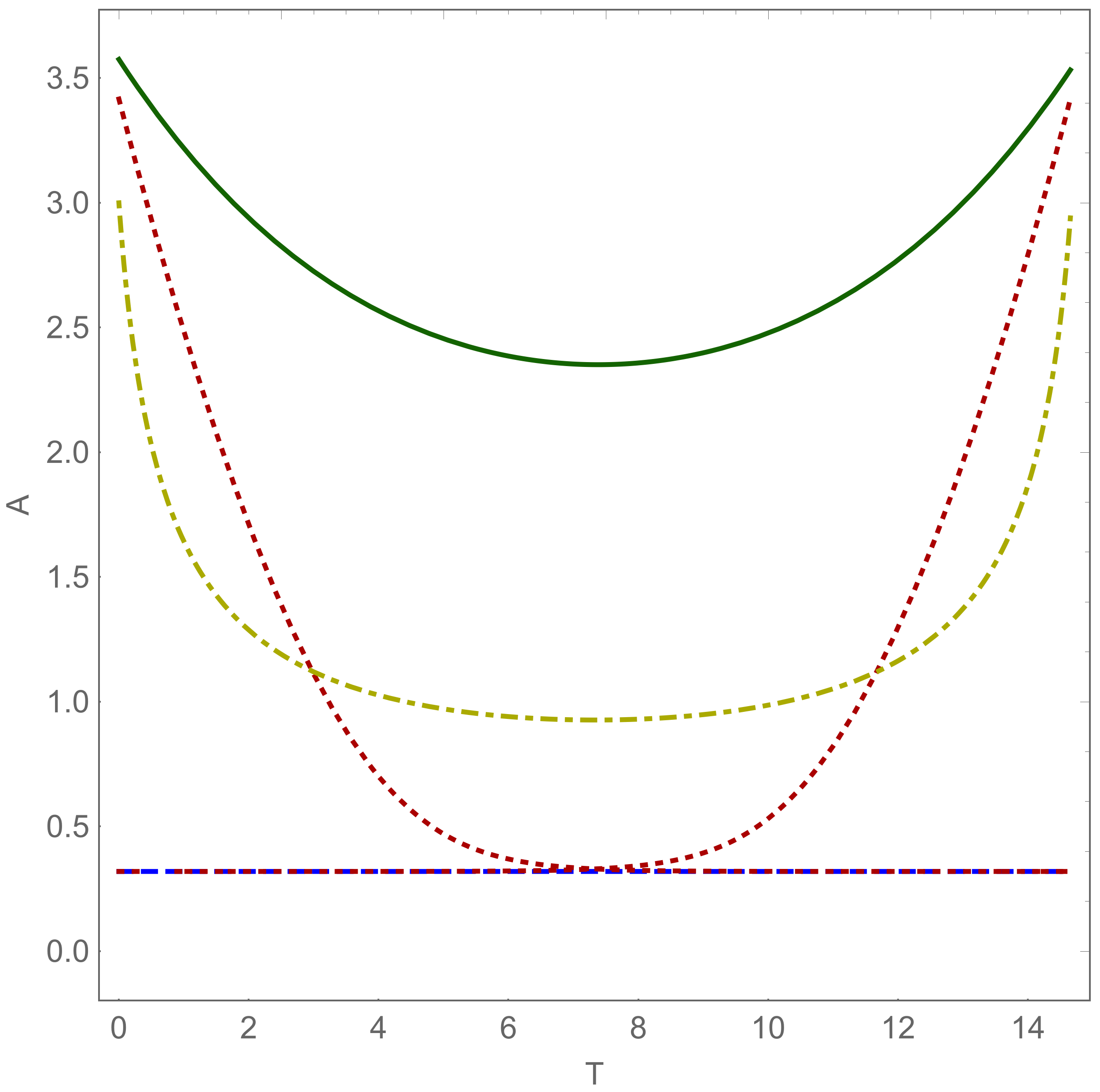}
		\caption{Bouncing trajectories compared to classical expansion away from and collapse toward the horizon, for $M=0.4$ and $\beta=5$.} 
		\label{fig:trajectories_bounce}
	\end{subfigure}
	\vspace{1em}
	\begin{subfigure}{0.45\textwidth}
		\centering
		\includegraphics[width=\textwidth]{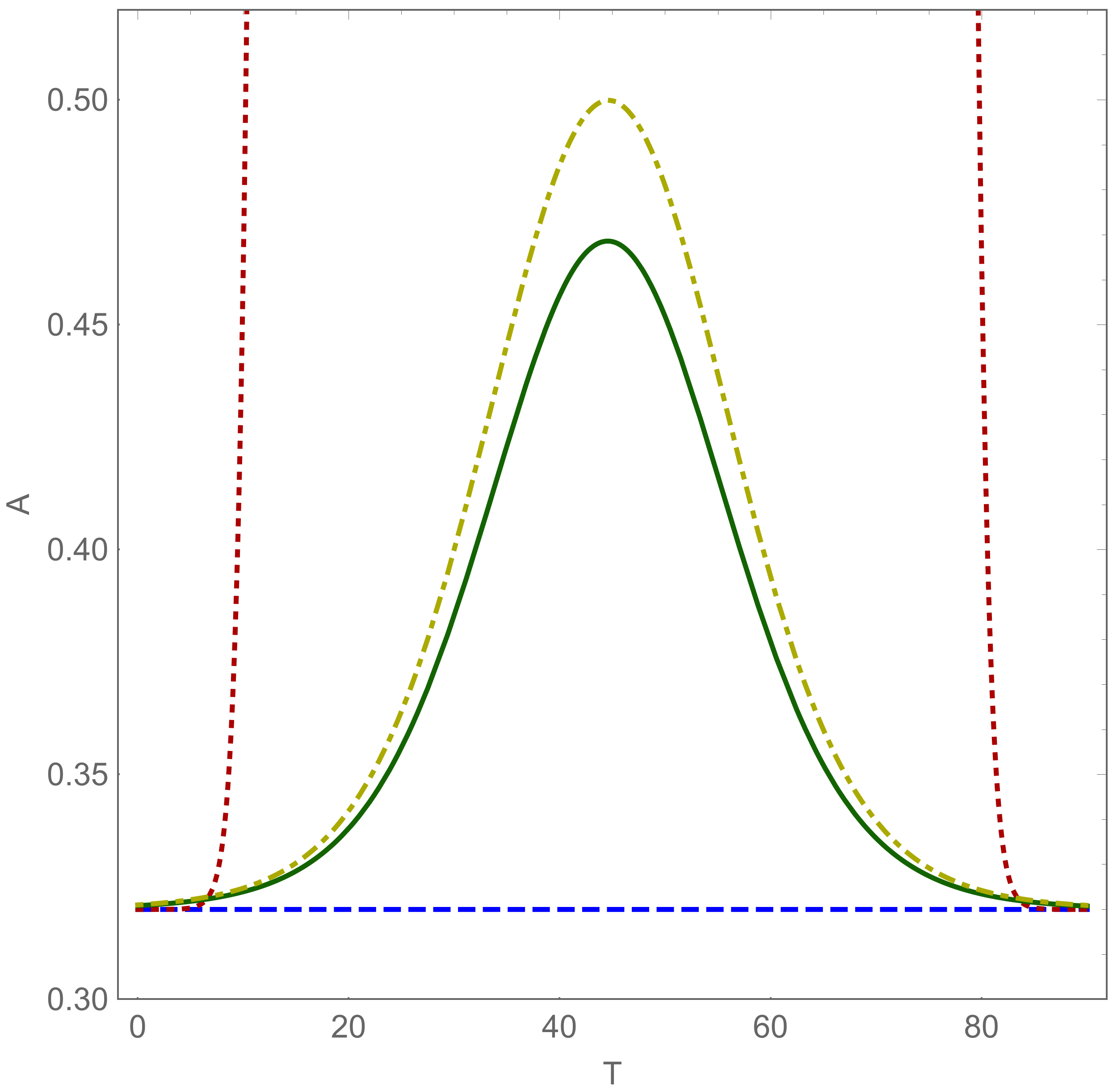}
		\caption{Recollapsing trajectories compared to classical expansion away from and collapse toward the horizon, for $M=0.4$ and $\beta=5$.} 
		\label{fig:trajectories_recollapse}
	\end{subfigure}
	\begin{subfigure}{0.45\textwidth}
		\centering
		\includegraphics[width=\textwidth]{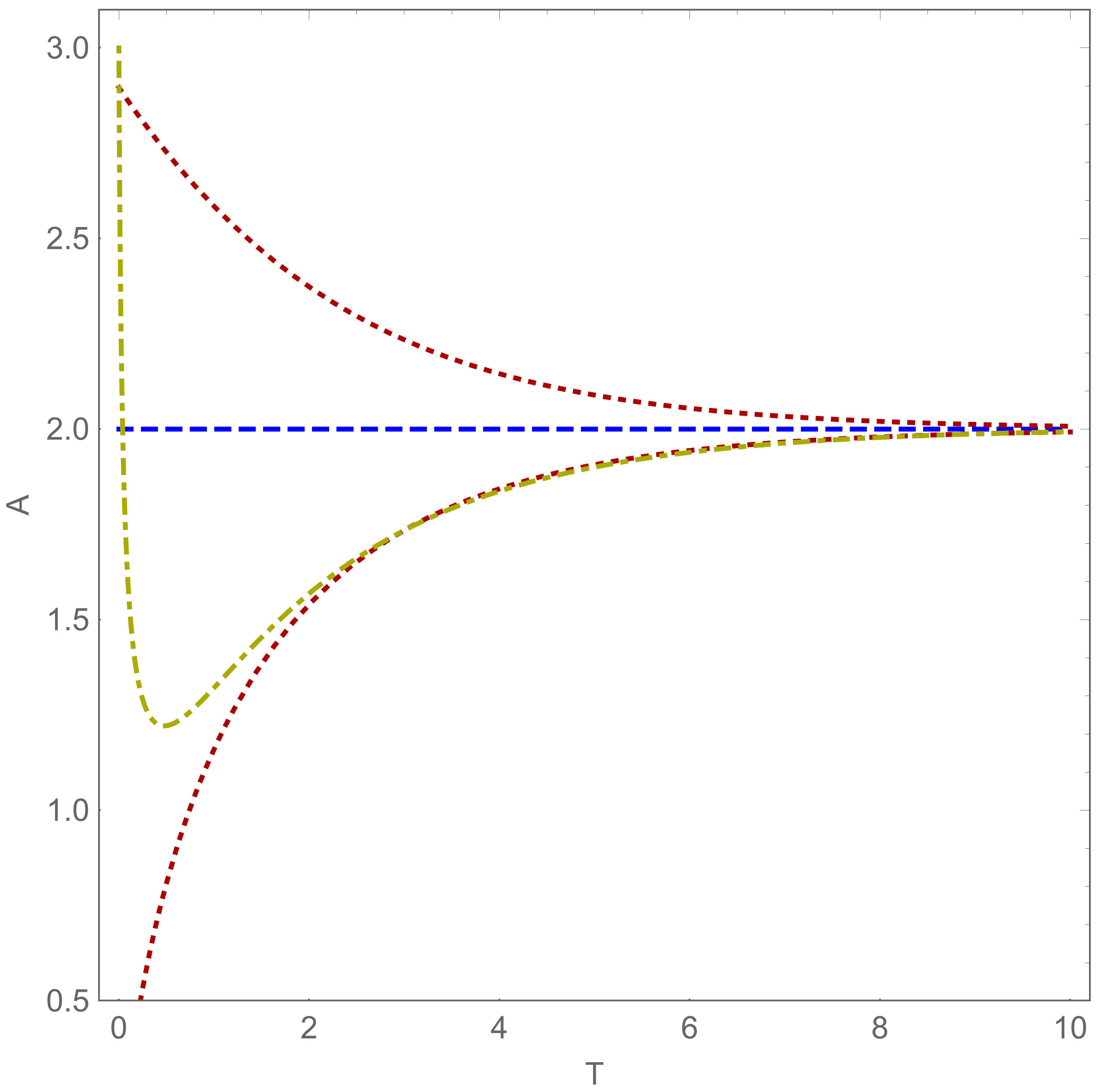}
		\caption{A trajectory collapsing from infinity and approaching the horizon from inside, compared to the classical approaches to the horizon both from inside and outside, for $M=1$ and $\beta=1$.} 
		\label{fig:trajectories_bounce_inside}
	\end{subfigure}
	\begin{subfigure}{0.45\textwidth}
		\centering
		\includegraphics[width=\textwidth]{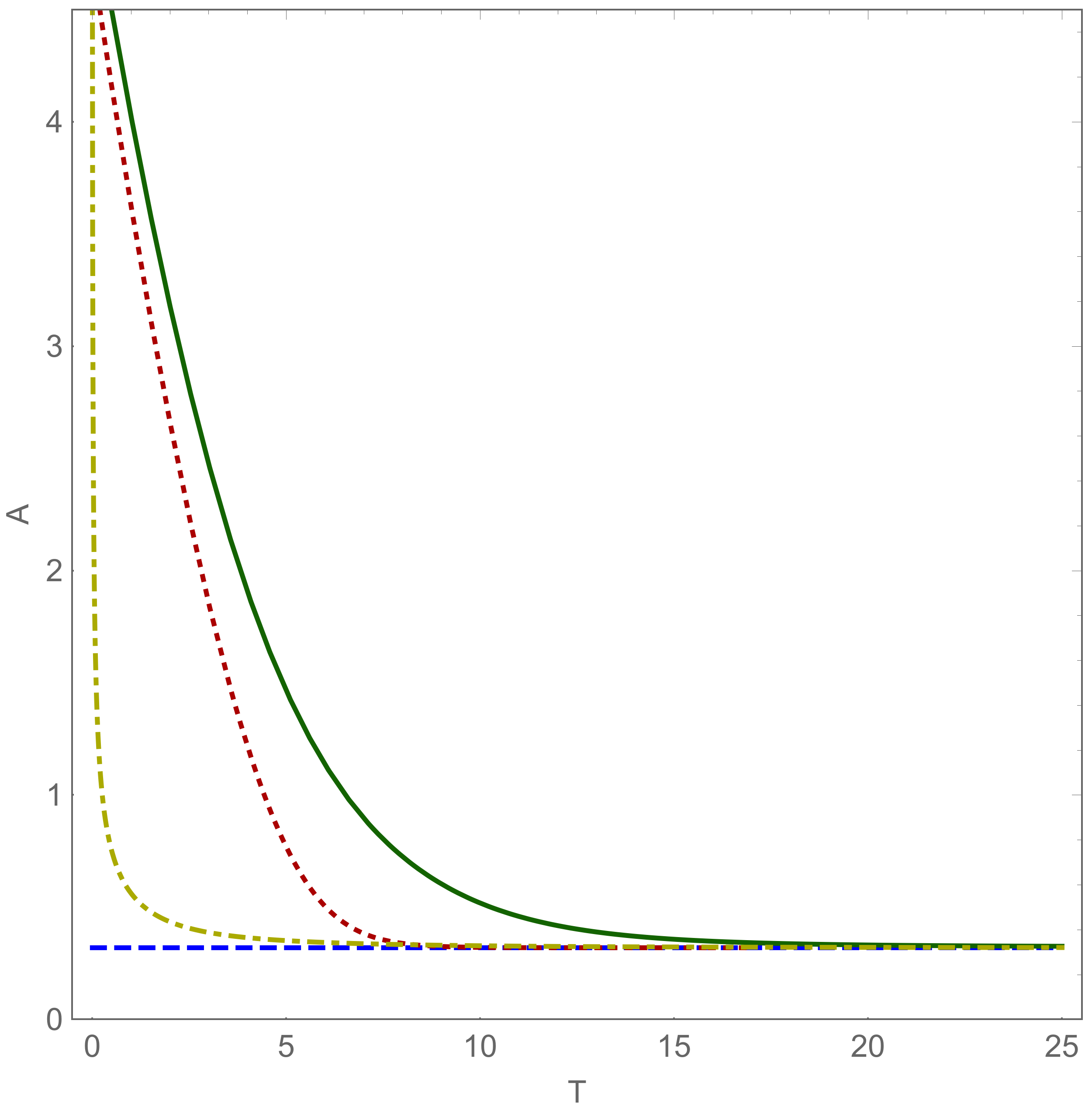}
		\caption{Trajectories asymptotically approaching the horizon from outside, compared to the classical counterpart, for $M=0.4$ and $\beta=1$.} 
		\label{fig:trajectories_asymptotic}
	\end{subfigure}
	\caption{The quantum corrected trajectories for the outside branch (full green lines) and inside branch (dotted dashed yellow lines) of the Hamiltonian \eqref{eq:hamiltonian_outinside} for different initial conditions as compared to their classical counterparts (dotted red lines) and the horizon $M=\sqrt{\frac{A}{2}}$ (dashed blue lines), in Planck units.}
	\label{fig:trajectories}
\end{figure*}

For lower $\beta$ as compared to $M$ the outside branch given by $\check{H}_+$ in Fig.\ \ref{fig:phase_space_portraits_outside} is qualitatively very close to the classical portrait; kinematically the system is split into two parts, asymptotically collapsing toward the horizon and expanding away from it. Solving the equations of motion shows that also dynamically the quantum corrected trajectory behaves similarly to the classical one, see Fig.\ \ref{fig:trajectories_asymptotic}. When one chooses a higher $\beta$, this picture changes dramatically: The branches collapsing from and expanding to infinity connect, as well as those near the horizon. This suggests a bounce of the dust cloud when collapsing from infinity, and a recollapse when expanding away from the horizon. This prediction is confirmed by the trajectories in Fig.\ \ref{fig:trajectories_bounce} and Fig.\ \ref{fig:trajectories_recollapse}. By comparison to the classical trajectories one can also see that the recollapse is a slower process than the bounce.

The inside branch given by $\check{H}_-$ behaves very similarly to the outside branch and is in sharp contrast to the classical case not confined to the inside of the horizon, see Fig.\ \ref{fig:phase_space_portraits_inside}. The only qualitative difference to the outside branch is that the asymptotic approach toward the horizon can also happen from the inside, after a bounce close to the classical singularity. For fixed $\beta$ this seems to take place for higher $M$ than the approach to the horizon from the outside. All of this can also be seen in the trajectories in Fig.\ \ref{fig:trajectories}.

We also want to note that these quantum corrected trajectories demonstrate that we do not need to consider explicitly constructing multivalued states at this stage. It seems that the branching points of the Hamiltonian $P\to\pm\infty$ can never be reached in finite time. 

\begin{figure*}
	\centering
	\begin{subfigure}{0.45\textwidth}
		\centering
		\includegraphics[width=\textwidth]{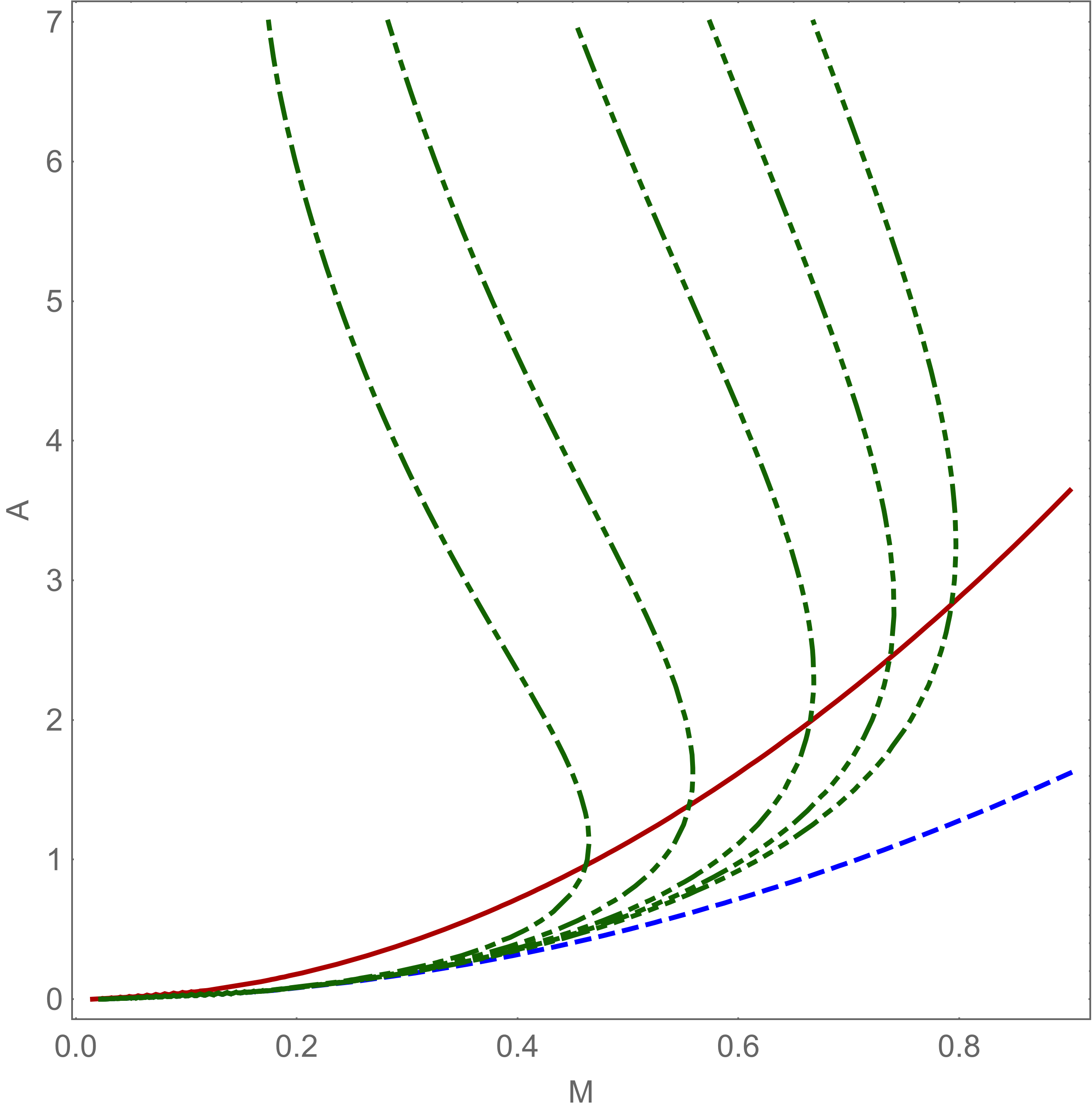}
		\caption{outside branch}
	\end{subfigure}
	\begin{subfigure}{0.45\textwidth}
		\centering
		\includegraphics[width=\textwidth]{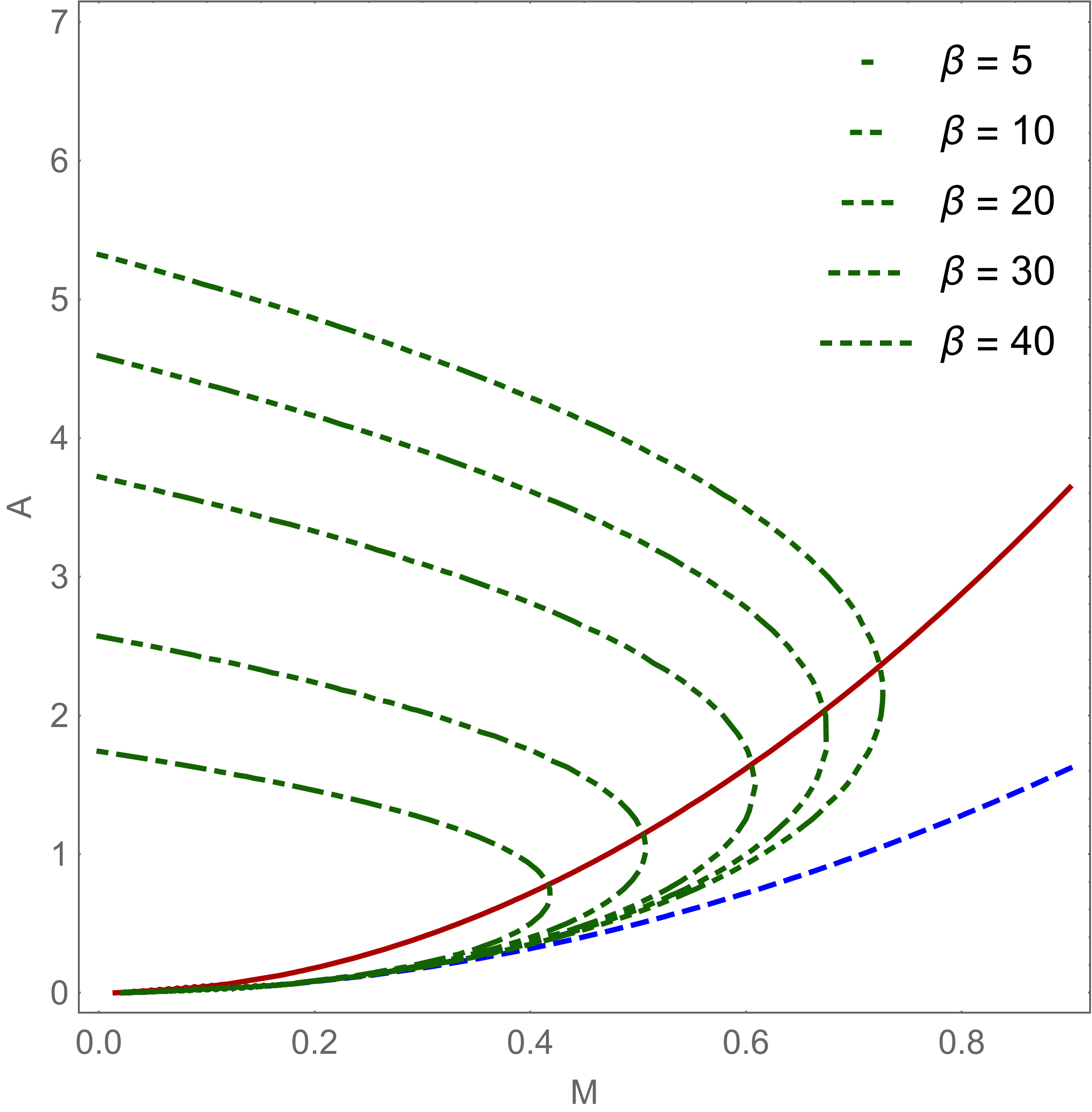}
		\caption{inside branch}
	\end{subfigure}
	\caption{The quantum corrected phase space portraits at $P_A=0$ for the outside and inside branch of the Hamiltonian as given by \eqref{eq:phase_space_portrait} for different $\beta$ (green lines) as compared to the horizon $M=\sqrt{\frac{A}{2}}$ (dashed blue line) and the photon sphere $M=\frac{\sqrt{2A}}{3}$ (full red line), in Planck units.}
	\label{fig:phase_space_p0}
\end{figure*}

Next we want to take a closer look at how this transition in behavior depends on $M$ and $\beta$. To this end we plot phase space portraits at $P_A=0$ as a function of $M$ for different $\beta$, see Fig.\ \ref{fig:phase_space_p0}. When for given $M$ and $\beta$ there are two values of $A$ on the phase space contour we have a bounce and a recollapse, and when there are none we have an asymptotic approach to the horizon.

With this in mind we see that for every $\beta$ all masses up to some critical mass bounce and recollapse, and all above it do not. This critical mass grows with increasing $\beta$. Aside from the fact that the minimal area of the bounce grows slower with decreasing $M$ for the inside branch than for the outside branch, there is no qualitative difference between the two branches for this aspect of the dynamics either.

In addition we can read off from Fig.\ \ref{fig:phase_space_p0}. that the minimal area of bouncing dust clouds lies outside of the photon sphere, and the maximal area of recollapsing ones between photon sphere and horizon. This is somewhat discouraging with regard to bouncing collapse: roughly speaking, everything outside of the photon sphere is visible to an outside observer, so these quantum corrected dynamics suggest that the dust cloud will never resemble anything even close to a black hole. Furthermore the minimal area grows with increasing $\beta$, while the maximal area decreases, suggesting that the details of the dynamics strongly depend on the fiducial vector.

The above suggests that if we want all dust clouds to bounce regardless of mass, or at least those with masses relevant for astrophysical considerations, we need to consider the case of very high $\beta$ relative to unity. For these values of $\beta$ and $M$ a numerical treatment seems to be quite challenging. Luckily we can estimate the integrals in $\check{H}_\pm$ for the case $\beta\to\infty$ by using a saddle point approximation, allowing us to check the behavior of the system analytically.

First we note that the factor in front of the Hamiltonians in \eqref{eq:hamiltonian_outinside} approaches $1$ when $\beta\to\infty$. Further we slightly rewrite the Hamiltonians,
\begin{multline}
\check{H}_\pm(P_A,A)=- \sqrt{\frac{A}{2}} +\frac{ \beta^{2\beta+\frac{1}{2}} }{2^\frac{5}{2}\hbar^2 \Gamma\!\left(2\beta +\tfrac{1}{2} \right) }\\\int_{0}^\infty dx\int_{0}^\infty dy~ \frac{ x - y }{ F_\pm\left(\frac{\pi }{2\hbar}\left(  x - y\right) \right) } \frac{\left( x+y\right)^{\frac{3}{2}}  }{ x y } \\e^{\frac{i}{\hbar}P_A\left(x-y \right) +\beta\left[- \frac{x+y}{A} + 2 \ln\left( \frac{4}{A} \frac{xy}{x+y} \right)   \right] }    .
\end{multline}
The function in the exponential multiplied by $\beta$ has a single critical point at $x=y=A$, where it takes the value $2\ln2-2$. The eigenvalues of its Hessian are $-1/A^2$ and $-2/A^2$, meaning the critical point is a maximum. The Hamiltonians for $\beta\to\infty$ can then be approximated as 
\begin{equation}
\check{H}_\pm(P_A,A)\sim- \sqrt{\frac{A}{2}} +  \sqrt{\frac{A^3}{\pi\beta\hbar^2 }}  ,
\end{equation}
where we have used that
\begin{equation}
\Gamma\!\left(a z+b \right)\sim \sqrt{2\pi} e^{-az} (az)^{az+b-\frac{1}{2}}
\end{equation}
for $a>0$ and $\beta\to\infty$ \cite{DLMF}, and
\begin{equation}
\left.\frac{x}{F_\pm\left(a x \right) }\right|_{x=0} = \frac{1}{a} .
\end{equation}
The end result is the same for both the inside and the outside branch. This matches our earlier observation that the inside and outside branches show similar behavior for higher $\beta$. The Hamiltonian furthermore does not depend on the momentum in this limit. Since we are mainly interested in the $P_A=0$ region of phase space this is no obstacle; based on the series of the complex exponential in the Hamiltonians for $x\to y$ we can deduce that the subleading order in the $\beta\to\infty$ expansion will depend linearly on $P_A$ and hence vanish for $P_A=0$, leaving us with the above.  

We are interested in real, positive solutions of
\begin{equation}
0= R - \frac{R^3}{\sqrt{2\pi\beta\hbar^2 }} -2M ,
\end{equation}
where we have replaced $A$ by the radius of the dust cloud, $A=\frac{1}{2}R^2$. The polynomial above has a minimum at $R_-=-\frac{1}{\sqrt{3}}(2\pi\beta\hbar^2)^\frac{1}{4}$ and a maximum at $R_+=\frac{1}{\sqrt{3}}(2\pi\beta\hbar^2)^\frac{1}{4}$. Depending on the value of the polynomial at these extrema we have either one, two or three solutions.

For $M>0$ the minimum can never be positive, but the maximum can be negative for $\sqrt{2\pi\beta\hbar^2}<27M^2$. In this case we have a single root with $R<R_-<0$, and hence no positive solution. This reflects what we have seen in the numerical investigation: for small $\beta$ as compared to $M$, the dust cloud does not bounce.

When $\sqrt{2\pi\beta\hbar^2}>27M^2$ the maximum is positive, and we have in addition to the negative root a positive root at $R>R_+>3M$, and one at $R_-<R<R_+<3M$. Since for $R=0$ the polynomial is negative, the last root has to be positive. It is easy to see that any positive root has to fulfill $R>2M$. This also agrees with our previous results; for higher $\beta$ the dust cloud bounces with a minimal radius outside of the photon sphere or recollapses with maximal radius between horizon and photon sphere, depending on the initial conditions. Furthermore, the maximal radius grows with increasing $\beta$ while the minimal radius shrinks towards the horizon.

At $\sqrt{2\pi\beta\hbar^2}=27M^2$ both positive roots then join at $R=R_+=3M$, the minimal radius of the bounce and the maximal radius of the recollapse meeting at the photon sphere. 

If one finally takes the limit $\beta\to\infty$, the minimal radius gets pushed out to infinity, and the maximal radius to the horizon. The only valid solutions are then dust clouds stuck at the horizon, implying that one can never choose a single $\beta$ such that dust clouds with arbitrary masses bounce. 

Note that the same results can be obtained for a different fiducial vector, see App.\ \ref{app:D}, demonstrating a certain robustness of the behavior described above with regard to the quantization ambiguities. 

Lastly we want to mention that the classical limit in ACSQ is not simply $\hbar\to0$, but also requires taking a limit in the family of fiducial vectors such that $|\Phi(x)|^2\to\delta(x-1)$. For our choice of fiducial vector this translates to $\beta\to\infty$, but obviously taking $\hbar\to0$ in the approximated Hamiltonian above does not yield the classical limit. This is because the two limits $\hbar\to0$ and $\beta\to\infty$ cannot simply be taken one after the other to get the classical limit, one rather has to let $\beta\to\infty$ as a function of $\hbar$ as $\hbar\to0$, as also discussed in Ref.\ \cite{BergeronBounce}. Details can be found in App.\ \ref{app:C}.


\section{Connecting the two observers} \label{ch:chapter_6}

In the last sections we have developed two different quantum theories for OS collapse, one for the comoving observer and one for the stationary exterior observer. When we assume that those two theories are different sectors in a full theory for a quantum OS model, fully covariant and thus incorporating every possible observer, it becomes necessary to find a way to connect the two specialized theories. Here we want to explore a possible such connection; we want to find a way to switch observers in the quantum theory.

To this end we first note that 
\begin{equation}
\left\{ R,\Pi_\pm\right\}=\left(1-\frac{\Pi_\pm^2}{R^2} \right) ,
\end{equation} 
where we define $\Pi_+=R\tanh\frac{P}{R}$ and $\Pi_-=R\coth\frac{P}{R}$. Hence we can express the Hamiltonian for the stationary observer \eqref{eq:hamiltonian_Killing_time} as
\begin{equation}
H_\text{eff} = - \frac{\Pi^2}{2R} , \label{eq:effective_H}
\end{equation}
with the modified Poisson bracket
\begin{equation}
\left\{ R,\Pi\right\}=\left(1-\frac{\Pi^2}{R^2} \right) .\label{eq:mod_bracket}
\end{equation}
It is apparent that the Hamiltonian is identical to the one for the comoving observer. The difference between the two classical deparametrized theories is then only in the Poisson bracket; for the comoving observer we have the usual $\left\lbrace R,P \right\rbrace=1 $, and for the stationary observer we have the above. It seems reasonable to assume that this observation could be generalized also to other time coordinates, corresponding to other observers. 

As a side note we want to mention that in this form both branches of the Hamiltonian are unified. The dynamics of the outside are recovered if we restrict to $|\Pi|<R$, and the ones of the inside for $|\Pi|>R$. It will become apparent later that difficulties with finding a similarly unified quantum representation hinders the construction of a quantum theory with both branches, once again only leaving the prescription with multivalued states to accomplish this. 

In the following we want to show that the above observation carries over from the classical theories to the quantum theories: we will demonstrate that the quantum theory corresponding to the stationary observer as discussed in Sec.\ \ref{ch:chapter_5} can also be considered as a quantization of \eqref{eq:effective_H} with the modified Poisson bracket \eqref{eq:mod_bracket}, reproduced in the commutator $[\hat{R},\hat{\Pi}]$.

Note that \eqref{eq:mod_bracket} also emerges, in a slightly modified form, in the context of noncommutative spaces for the so called Snyder models. In fact we can exactly reproduce the modified bracket corresponding to the anti-Snyder model by using $A=\frac{1}{2}R^2$. For a discussion of these models, including how they can be quantized, see Ref.\ \cite{MignemiSnyderSpace} and references therein. Here we will employ ACSQ to fulfill these commutation relations.

It turns out that it is useful to follow what we did classically; there we promoted the functions $\Pi_\pm(P,R)$ to the phase space coordinate $\Pi$, so let us proceed analogously for their operators.

Let us first consider the operator associated with $\Pi_+$ using the quantization map from previous sections,
\begin{align}
\hat{\Pi}_+\psi(x)&=\frac{1}{2\pi\hbar c_{-1}^\Phi}\int_{0}^{\infty}dR\int_{-\infty}^{\infty}dP\int_{0}^{\infty}dy~\tanh\frac{P}{R}\\&\quad\quad e^{\frac{i}{\hbar}(x-y)P}\Phi\left(\tfrac{x}{R}\right)\Phi^*\left(\tfrac{y}{R}\right)~\psi(y)\\
&\hspace{-1.8em}\overset{\Pi=R\tanh\frac{P}{R}}{=} \frac{1}{2\pi\hbar c_{-1}^\Phi}\int_{0}^{\infty}dR\int_{-R}^{R}\frac{d\Pi}{1-\frac{\Pi^2}{R^2}}\int_{0}^{\infty}dy~\frac{\Pi}{R} \\&\quad\quad e^{\frac{i}{\hbar}(x-y)\frac{R}{2}\ln\left(\frac{1+\frac{\Pi}{R}}{1-\frac{\Pi}{R}} \right) }\Phi\left(\tfrac{x}{R}\right)\Phi^*\left(\tfrac{y}{R}\right)~\psi(y) ,
\end{align}
where we have expressed the inverse hyperbolic tangent in terms of the logarithm. The last expression above looks suspiciously like an ACSQ of $\Pi$ in a new parametrization of the affine group by the phase space variables as discussed in \cite{MeParameterizations}. Along the same line we can proceed for $\Pi_-$,
\begin{align}
\hat{\Pi}_-\psi(x)&=\frac{1}{2\pi\hbar c_{-1}^\Phi}\int_{0}^{\infty}dR\int_{-\infty}^{\infty}dP\int_{0}^{\infty}dy~\coth\frac{P}{R}\\&\quad\quad e^{\frac{i}{\hbar}(x-y)P}\Phi\left(\tfrac{x}{R}\right)\Phi^*\left(\tfrac{y}{R}\right)~\psi(y)\\
&\hspace{-1.8em}\overset{\Pi=R\coth\frac{P}{R}}{=} \frac{1}{2\pi\hbar c_{-1}^\Phi}\int_{0}^{\infty}dR\int_{\mathbb{R}\backslash [-R,R]} \frac{d\Pi}{\frac{\Pi^2}{R^2}-1}\int_{0}^{\infty}dy\\&\quad\quad \frac{\Pi}{R}~ e^{\frac{i}{\hbar}(x-y)\frac{R}{2}\ln\left(\frac{\frac{\Pi}{R}+1}{\frac{\Pi}{R}-1} \right) }\Phi\left(\tfrac{x}{R}\right)\Phi^*\left(\tfrac{y}{R}\right)~\psi(y) .
\end{align}
The above suggests using the quantization maps
\begin{multline}
\hat{f} \psi(x) = \frac{1}{2\pi\hbar c_{-1}^\Phi}\int_{0}^{\infty}\frac{dR}{R}\int_I\frac{d\Pi}{\left| 1-\frac{\Pi^2}{R^2}\right| }\int_{0}^{\infty}dy~f(R,\Pi)\\e^{\frac{i}{\hbar}(x-y)\frac{R}{2}\ln\left|\frac{1+\frac{\Pi}{R}}{1-\frac{\Pi}{R}} \right| }\Phi\left(\tfrac{x}{R}\right)\Phi^*\left(\tfrac{y}{R}\right)~\psi(y) ,\label{eq:new_quant}
\end{multline}
assigning $\hat{f}$ to the phase space function $f(\Pi,R)$, where 
\begin{equation}
I=\left[-R,R \right]\quad\text{or}\quad I=\mathbb{R}\backslash [-R,R] .
\end{equation}

The above is indeed equivalent to ACSQ with an alternative parametrization of the affine group by
\begin{equation}
R\in\mathbb{R}_+\,,\quad\frac{R}{2}\ln\left|\frac{1+\frac{\Pi}{R}}{1-\frac{\Pi}{R}} \right|\in\mathbb{R} ,\label{eq:new_para}
\end{equation}
leading to coherent states
\begin{equation}
\braket{x|\Pi,R}=\frac{1}{\sqrt{R}}e^{\frac{i}{\hbar}x \frac{R}{2}\ln\left|\frac{1+\frac{\Pi}{R}}{1-\frac{\Pi}{R}} \right|}\psi\left( \tfrac{x}{R}\right) .
\end{equation}

That these states indeed lead to a resolution of the identity with the measure in \eqref{eq:new_quant} directly follows from the fact that we constructed them by substitution from the coherent states used in the previous sections. The quantization map \eqref{eq:new_quant} is therefore well defined. For details see Ref.\ \cite{MeParameterizations}.

We necessarily need to restrict $\Pi$ either to $|\Pi|>R$ or $|\Pi|<R$ in order for the parametrization above to cover the affine group only once. This is what we alluded to earlier: at this point the inside and outside branch once again split up.

As one can see from how we constructed the quantization maps \eqref{eq:new_quant}, the operator associated with a phase space function $f(R,\Pi)$ will be equivalent to one of two operators quantized according to the ACSQ prescription we have used in the previous chapters, one associated with $f(\Pi_+(P,R),R)$ and one with $f(\Pi_-(P,R),R)$, depending on which of the maps we are using. It is then easy to see that the operator corresponding to the Hamiltonian \eqref{eq:effective_H} in these new quantization maps is equivalent to one of the two branches of \eqref{eq:hamiltonian_Killing_time} quantized with the original quantization map.

Let us now demonstrate that these quantization maps indeed reproduce \eqref{eq:mod_bracket}. $\hat{R}$ still acts as a multiplication operator,
\begin{equation}
\hat{R}\psi(x)=\frac{ c_{0}^\Phi}{ c_{-1}^\Phi}\,x\,\psi(x) .
\end{equation}
We can then write
\begin{align}
\hspace{-2em}[\hat{R},&\hat{\Pi}]=\frac{c_{0}^\Phi}{2\pi\hbar {c_{-1}^\Phi}^2}\int_{0}^{\infty}\frac{dR}{R}\int_I\frac{d\Pi}{\left| 1-\frac{\Pi^2}{R^2}\right| }\int_{0}^{\infty}dy~(x-y)\\&\quad\quad~\Pi~e^{\frac{i}{\hbar}(x-y)\frac{R}{2}\ln\left|\frac{1+\frac{\Pi}{R}}{1-\frac{\Pi}{R}} \right| }\Phi\left(\tfrac{x}{R}\right)\Phi^*\left(\tfrac{y}{R}\right)~\psi(y)\\
&=-\frac{i c_{0}^\Phi}{2\pi {c_{-1}^\Phi}^2}\int_{0}^{\infty}\frac{dR}{R}\int_I d\Pi\int_{0}^{\infty}dy~\sgn\left(1-\frac{\Pi^2}{R^2}\right)\\&\quad\quad\Pi~\frac{\partial}{\partial\Pi}e^{\frac{i}{\hbar}(x-y)\frac{R}{2}\ln\left|\frac{1+\frac{\Pi}{R}}{1-\frac{\Pi}{R}} \right| }\Phi\left(\tfrac{x}{R}\right)\Phi^*\left(\tfrac{y}{R}\right)~\psi(y)\\
&=\frac{i c_{0}^\Phi}{2\pi {c_{-1}^\Phi}^2}\int_{0}^{\infty}\frac{dR}{R}\int_I \frac{d\Pi}{\left| 1-\frac{\Pi^2}{R^2}\right| }\int_{0}^{\infty}dy~\left(1-\frac{\Pi^2}{R^2}\right)\\&\quad\quad e^{\frac{i}{\hbar}(x-y)\frac{R}{2}\ln\left|\frac{1+\frac{\Pi}{R}}{1-\frac{\Pi}{R}} \right| }\Phi\left(\tfrac{x}{R}\right)\Phi^*\left(\tfrac{y}{R}\right)~\psi(y)\\
&=i\hbar\frac{c_{0}^\Phi}{c_{-1}^\Phi}\widehat{\left\{ R,\Pi\right\}} .
\end{align}
Note that to suppress any boundary terms that arise from the partial integration we need make use of the usual regularization of Fourier transformations for $\Pi\to\pm R$, corresponding to the momentum in the Fourier transformation going to $\pm\infty$. If we choose the inside branch of the quantization map we also need to suppress boundary terms at $\Pi\to\pm\infty$, corresponding to the regularization of $\coth(x)$ at $x=0$ we made use of in the last section.

The calculation above further suggests that one can easily implement other modified Poisson brackets in ACSQ by replacing the momentum in our original quantization by the integral of the right hand side of the modified bracket over the new momentum, as long as the resulting function covers the real line at least once. Aside from leading further support to the feasibility of generalizing this procedure to different observers, this could be of use when studying e.g.\ the aforementioned Snyder spaces, or theories with Dirac brackets.

At this point we can conclude that the quantization map works as designed: It reproduces the modified Poisson brackets \eqref{eq:mod_bracket} and maps \eqref{eq:effective_H} to the Hamilton operators used in Sec.\ \ref{ch:chapter_5}. Our quantization in Sec.\ \ref{ch:chapter_5} also made use of a simple canonical transformation, but as already argued in Ref.\ \cite{KlauderCanonical} this does not affect the underlying quantum theory. In App.\ \ref{app:E} we explicitly demonstrate that also the quantum corrected considerations emerge unscathed from the new quantum theory: the modification of the Poisson bracket carries over as is to the quantum corrected equations. 

We have thus demonstrated that the reduced quantum theories from Secs.\ \ref{ch:chapter_4} and \ref{ch:chapter_5} can be related by modifying the commutation relations. In ACSQ this is further equivalent to switching between parametrizations of the affine group in the quantization map.

By referencing the quantization map we still involve the classical theory, but conceptually it would be more satisfactory to discuss the relation between the quantum theories purely at the quantum level. This is possible by considering what we discussed above as a map $\mathcal{T}$ on the operator algebra. Lastly we want to share a few observations about this map.

We know how it acts at least on operators corresponding to phase space functions $f(P,R)$,
\begin{equation}
\mathcal{T}( \hat{f} ) = \hat{f}_\pm ,
\end{equation}
where $f_\pm(P,R)=f(\Pi_\pm(P,R),R)$, and hats exclusively denote quantization via the original map from now on.

$\mathcal{T}$ has to be linear because the quantization map is, and it has to be bijective, at least when restricted to operators corresponding to phase space functions. Furthermore we know that it cannot be an isomorphism: it changes the commutation relations by construction but leaves the identity invariant:
\begin{equation}
\mathcal{T}( [\hat{R},\hat{P} ] )=i\hbar\,\mathcal{T}(\mathds{1} )= i\hbar\,\mathds{1} \neq  [\mathcal{T}(\hat{R}),\mathcal{T}(\hat{P}) ] = [\hat{R},\hat{\Pi} ] .
\end{equation}

This map is furthermore not equivalent to any transformation, including non-linear ones, acting on the Hilbert space $\mathcal{H}$. This can be seen as follows. Such an equivalent transformation $\bar{\mathcal{T}}$ would be defined by
\begin{equation}
\braket{\psi,\mathcal{T}(\hat{O})\chi } = \braket{\bar{\mathcal{T}}(\psi),\hat{O}\bar{\mathcal{T}}(\chi) }
\end{equation}
for all $\psi,\chi\in\mathcal{H}$ and all operators $\hat{O}$. For clarity we switch from Dirac notation to explicitly writing out the scalar product on $\mathcal{H}$.
The above has to hold in particular for the identity operator $\mathds{1}=\mathcal{T}\left(\mathds{1} \right)$, such that
\begin{equation}
\braket{\psi,\chi } = \braket{\bar{\mathcal{T}}(\psi),\bar{\mathcal{T}}(\chi) }
\end{equation}
It is then easy to see that $\bar{\mathcal{T}}$ has to be linear. Since linear operators have an adjoint, one can then find from the above that $\bar{\mathcal{T}}$ further has to be unitary. As follows directly from Ref.\ \cite{MeParameterizations}, this is not possible. Even without reference to Ref.\ \cite{MeParameterizations} it leads to a contradiction, because it would imply that $\mathcal{T}$ is an isomorphism of the operator algebra,
\begin{align}
\braket{\psi,\mathcal{T}([\hat{O}_1,\hat{O}_2 ])\,\chi } &=	\braket{\bar{\mathcal{T}}\psi,[\hat{O}_1,\hat{O}_2 ]\,\bar{\mathcal{T}}\chi}\\
&=\braket{\psi,\bar{\mathcal{T}}^\dagger[\hat{O}_1,\hat{O}_2 ]\bar{\mathcal{T}}\,\chi}\\
&=\braket{\psi,[\bar{\mathcal{T}}^\dagger\hat{O}_1\bar{\mathcal{T}},\bar{\mathcal{T}}^\dagger\hat{O}_2\bar{\mathcal{T}} ]\,\bar{\mathcal{T}}\chi}\\ &=\braket{\psi,[\mathcal{T}( \hat{O}_1) ,\mathcal{T}( \hat{O}_2) ]\,\chi} .
\end{align}

We can thus say that the switch between the two observers happens not at the level of the Hilbert space, but of the operator algebra and in particular its interpretation in terms of physical quantities. We want to stress that the above rests on the fact that the modification of the commutation relation is built into the quantization map. Dropping this feature and sticking to the two different Hamiltonians we used originally, there might exist a transformation on $\mathcal{H}$ that maps the two Hamiltonians onto each other. For the explicit construction of such a transformation for a different system see Ref.\ \cite{VanrietveldeRelational}.  

In conclusion we can say that our two quantum theories, each corresponding to one of the observers, can be related by changing the quantization map in such a way that the canonical commutation relations are modified. This relation can not be represented as a transformation between Hilbert spaces, and is then in particular not unitary.

This is unfortunate, since one generally expects states representing different observer in a full quantum gravity theory to be connected by unitary transformations. We can offer in defense of our construction that at the level of such deparametrized theories as we discuss here a unitary relation between observers might not be strictly necessary, as long as predictions made still show some form of consistency between observers.

Here we have only demonstrated a very basic such consistency: we have shown that both observers see a bounce when the quantization ambiguities are chosen accordingly. Before we can really say with confidence that what we discussed here could be developed into a candidate for a notion of quantum covariance, the predictions made for each observer need to be compared more carefully.

The most obvious way forward in this regard seems to be an application of this idea to more straightforward coordinate transformation than we have considered here. Beyond its elaborate functional form, the Painlev\'{e}-Gullstrand transformation also brings with it the further conceptual complication of splitting up the system into inside and outside of the horizon. It might thus be interesting to consider models collapsing to a naked singularity, such as the Lema\^{i}tre-Tolman-Bondi model, to circumvent the latter problem.

Furthermore, leaving the realm of black holes and focusing on time-reparametrization invariant systems might simplify matters further. Applying our construction to quantum cosmology proper is not straightforward, since there Hamiltonian constraints are generally not deparametrizable. However, one could as a first step discuss only cosmological models generated by Brown-Kucha\v{r} dust. We leave this for future work.


\section{Conclusions} \label{ch:chapter_7}

In this paper we have discussed a quantum OS model using ACSQ based on our previous classical considerations in Ref.\ \cite{MeOS}. More specifically we have quantized two Hamiltonians describing the flat OS model from the point of view of two observers, one comoving with the dust and one stationary outside of it.

Especially quantizing the latter Hamiltonian has nicely illustrated the benefits of ACSQ. In Dirac quantization this Hamiltonian does not even have a well defined operator due to the occurrence of hyperbolic functions and square roots. In ACSQ every phase space function can at least formally be identified with an operator, and if the function is at least semi-bounded we can even be sure that it has a self-adjoint extension. This is particularly useful here, because the Hamiltonian operator turns out to be a complicated integral operator for which it is difficult to find eigenfunctions. One can further find quantum corrected dynamics of the system without explicitly considering the Hamilton operator by working with the lower symbol of the Hamiltonian.

We want to emphasize again that while usually these quantum corrected dynamics are interpreted as exact in their own semiclassical theory \cite{KlauderEnhanced}, we have demonstrated here that they can also be understood within the full quantum theory; they follow from the Schr\"odinger equation as approximate dynamics of coherent states.

Of course if one wants to go beyond quantum corrections, directly working with the operator cannot be helped. In particular determining how accurate the semiclassical approximation is depends on the stability of the coherent states. Determining this stability seems to be a problem close in complexity to actually solving the Schr\"odinger equation. Nevertheless, using the lower symbol of the Hamiltonian is a convenient way to find quantum corrected dynamics of a system. 

The first Hamiltonian, describing the model as seen by a comoving observer, had already been quantized using Dirac's prescription for canonical quantization in the context of the Lema\^{i}tre-Tolman-Bondi model in Ref.\ \cite{MeLTB}. Here we have shown that for this particular Hamiltonian ACSQ and Dirac quantization lead to the same Hamilton operator. Further the quantum corrected dynamics found in ACSQ exactly match those found by investigating a wave packet in Dirac quantization in Ref.\ \cite{MeLTB}. These show that the dust cloud as seen by the comoving observer with quantum corrections bounces at a minimal radius scaling inversely with $M^\frac{1}{3}$, where $M$ is the total mass of the dust cloud.

For the second Hamiltonian, corresponding to the exterior stationary observer, the quantum corrected dynamics predict that the dust cloud as seen by the exterior observer can also bounce, where the minimal radius also grows larger with decreasing $M$. This is encouraging since it establishes a rudimentary consistency for our approach of switching observers in the classical canonical theory as described in Ref.\ \cite{MeOS}. Complementary to the bounce there emerges also a recollapse of dust clouds starting close to the horizon and expanding away from it. These dust clouds reach a maximal radius and then start approaching the horizon again.

The other details of the dynamics are unfortunately less satisfactory. The bounce is far less robust under quantization ambiguities than for the comoving observer. For any given mass the parameter $\beta$ in the family of fiducial vectors we used had to be chosen higher than a critical value $\propto M^4$ for a bounce to occur. For smaller $\beta$ the quantum corrected dynamics are qualitatively identical to the classical asymptotic approach to the horizon. As a consequence one can never choose a $\beta$ such that the dust cloud bounces for all $M$.

A feature of the bounce that we have shown to be quite robust with regard to quantization ambiguities is the fact that when the bounce occurs, the minimal radius is outside of the photon sphere $R=3M$. As seen by the exterior observer, the horizon never forms and the collapse never reaches a stage that even remotely resembles a black hole. The notion of a lifetime can then not even be defined. In a way, this bounce evades most of the conceptual problems of bouncing collapse somewhat elegantly by avoiding the horizon altogether. Thereby it unfortunately directly contradicts observations.

As a consequence we also have to concede that the ad hoc construction to find the lifetime from the point of view of the exterior observer in \cite{MeLTB}, leading to a lifetime $\propto M^3$, does not agree with the more rigorous implementation of this observer here. It is possible that this lifetime could be recovered when one goes beyond quantum corrections since it relies to some extent on `discrete' states; maybe transitions between our semiclassical states and states within the horizon are allowed in the theory but do not emerge in the quantum corrected dynamics. However, as mentioned above the unwieldy form of the Hamilton operator makes such investigations quite demanding.
 
We could now start speculating that the inclusion of pre-Hawking radiation could save the model, after all it has been shown to also affect classical collapse quite drastically \cite{MersiniHorizonAvoidance}. However, in hindsight our results are not very surprising. Our canonical formulation fixes the exterior of the dust cloud to be exactly Schwarzschild, and imposes the Painlev\'{e}-Gullstrand coordinate transformation between our two observers. Under these conditions avoiding the horizon altogether from the point of view of the exterior observer seems to be the only straightforward way to implement a bounce consistently between the two observers. While the matching conditions to the Schwarzschild exterior are not implemented explicitly, they influence our results implicitly in this way through the Painlev\'{e}-Gullstrand transformation. However, the deeper meaning of the photon sphere as the minimal radius remains unclear to us.

Note that the same line of reasoning does not apply to the comoving observer, for which why we did not find a lower bound for the minimal radius. To find the Hamiltonian generating the corresponding dynamics the Schwarzschild exterior is not relevant at all, it is identical to that of a Friedmann model.

A similar conclusion was drawn in Refs.\ \cite{AchourBounce1, AchourBounce2, AchourBounce3}, where the matching conditions between an effective bouncing interior and an arbitrary exterior was investigated. It emerged that the surface of the collapsing body must be in an untrapped region of the exterior at the time of the bounce. As already mentioned at the end of Sec.\ \ref{ch:chapter_2}, it seems to us that this procedure of finding an exterior to an effective bouncing interior through  matching conditions is a very promising approach to constructing a bouncing collapse model that is both conceptually and phenomenologically consistent, as was demonstrated in the aforementioned references. Coincidentally, it potentially makes it possible to also incorporate Hawking radiation according to Ref.\ \cite{MersiniHorizonAvoidance}.

Summarizing the parts of our paper relevant for bouncing collapse we can say that crucial ingredients seem to be missing from the program, and unfortunately this paper does not do much to fill in the blanks. These negative results however suggest where to look next, and we still believe that explicitly investigating the two distinguished observers in the OS model is a promising approach.

Lastly we want to comment on some aspects of this paper going beyond bouncing collapse. The Hamiltonian relevant for the stationary observer is multivalued and consists of two branches, one describing the behavior of the dust cloud outside of the horizon and one inside. In principle one would have to implement this Hamiltonian in the quantum theory by also considering multivalued states, but for our purposes this has turned out to be irrelevant: the quantum corrected trajectories never make use of the freedom to switch between the two branches, which allowed us to ignore this aspect for the discussion above. The results are the same as when we would have ignored the inside branch from the beginning.

The quantum corrected dynamics of the inside branch are interesting insofar as they behave very similarly to their outside counterpart. For $M$ and $\beta$ chosen accordingly there is a bounce and a recollapse outside of the horizon also for the inside branch. Also the asymptotic approach to the horizon can be obtained. The only qualitative difference between the two branches is that one can choose $M$ and $\beta$ in such a way that the collapse continues through the horizon and a bounce occurs close to the singularity followed by an asymptotic approach to the horizon from the inside.

Why the two branches of the quantum corrected Hamiltonian would behave so similarly when their classical counterparts occupy completely disjunct regions of phase space is puzzling to us. Because the most drastic departures of the quantum corrected inside branch from the classical trajectories occur close to $P_A=0$, where the classical Hamiltonian diverges, we cannot exclude the possibility that our regularization of this branch in Sec.\ \ref{ch:chapter_5} has led to this unintuitive behavior. 

However, since it is not necessary to consider both branches to find at least an internally consistent picture for the bounce, contradicting indications we found in Ref.\ \cite{MeOS}, this is not of primary importance. The disagreement with Ref.\ \cite{MeOS} in this respect is in our view not problematic, but rather a cautionary tale that excessively reformulating the classical constraints without any physical motivation can lead to wildly different quantum theories.

We have also observed that our two reduced quantum theories can be related by modifying the canonical commutation relation. The switch between observers at the level of these quantum theories then takes place on the operator algebra and its intepretation as physical quantities. It is impossible to implement it as a transformation on the Hilbert space.

It would be interesting to see whether this construction can be generalized beyond the present specialized case. If it can, another feature of ACSQ could become relevant; it seems to be relatively simple to implement at least some modified commutation relations by changing the parametrization of the affine group by the phase space. This is also of some interest for the quantization of other theories with modified Poisson brackets or Dirac brackets, see e.g.\ Refs.\ \cite{LiuMultivalued,RuzMultivalued,MignemiSnyderSpace}.


\section*{Acknowledgments}

The authors would like to thank Claus Kiefer and Roberto Casadio for valuable discussions and feedback on the manuscript, and are grateful to anonymous referees for constructive criticism. TS is grateful to Nick Kwidzinski, Yi-Fan Wang and Nick Nussbaum for further helpful discussions. This work was supported by the German-Polish bilateral project DAAD and MNiSW, No 57391638.

\vfill

\appendix


\section{Operators to elementary phase space functions}\label{app:0}

Here we want to investigate the elementary phase space functions $R$, $P$ and $D=RP$ in the context of ACSQ.

Consider first how $\hat{R}$, $\hat{P}$ and $\hat{D}$ act on wave functions,
\begin{align}
\hat{R}\psi(x)&=\frac{1}{2\pi\hbar c_{-1}^\Phi}\int_{0}^{\infty}dR\int_{-\infty}^{\infty}dP\int_{0}^{\infty}dy~e^{\frac{i}{\hbar}(x-y)P}\\&\quad\quad\Phi\!\left(\tfrac{x}{R}\right)\Phi^*\!\left(\tfrac{y}{R}\right)~\psi(y)\\
&=\frac{1}{ c_{-1}^\Phi}\int_{0}^{\infty}dR~\left| \Phi\!\left(\tfrac{x}{R}\right)\right| ^2 ~\psi(x)=\frac{ c_{0}^\Phi}{ c_{-1}^\Phi}\,x\,\psi(x),
\end{align}
\begin{align}
\hat{P}\psi(x)&=\frac{1}{2\pi\hbar c_{-1}^\Phi}\int_{0}^{\infty}dR\int_{-\infty}^{\infty}dP\int_{0}^{\infty}dy~\frac{P}{R}~e^{\frac{i}{\hbar}(x-y)P}\\&\quad\quad\Phi\!\left(\tfrac{x}{R}\right)\Phi^*\!\left(\tfrac{y}{R}\right)~\psi(y)\\
&=-\frac{i\hbar}{ c_{-1}^\Phi}\int_{0}^{\infty}\frac{dR}{R}~\Phi\!\left(\tfrac{x}{R}\right)\frac{\partial}{\partial x}~ \Phi^*\!\left(\tfrac{x}{R}\right)\psi(x)\\
&\overset{u=\frac{x}{R}}{=}-i\hbar\psi(x)'-\frac{i\hbar\psi(x)}{c^\Phi_{-1} x}\int_0^\infty du~\Phi(u)\Phi^*(u)'\\
&\equiv-i\hbar\psi(x)'-i\hbar\gamma\frac{\psi(x)}{x} ,
\end{align}
\begin{align}
\hat{D}\psi(x)&=\frac{1}{2\pi\hbar c_{-1}^\Phi}\int_{0}^{\infty}dR\int_{-\infty}^{\infty}dP\int_{0}^{\infty}dy~P~e^{\frac{i}{\hbar}(x-y)P}\\&\quad\quad\Phi\!\left(\tfrac{x}{R}\right)\Phi^*\!\left(\tfrac{y}{R}\right)~\psi(y)\\
&=-\frac{i\hbar}{ c_{-1}^\Phi}\int_{0}^{\infty}dR~\Phi\!\left(\tfrac{x}{R}\right)\frac{\partial}{\partial x}~ \Phi^*\!\left(\tfrac{x}{R}\right)\psi(x)\\
&\hspace{-0.6em}\overset{u=\frac{x}{R}}{=}-i\hbar\frac{c^\Phi_0}{c^\Phi_{-1}}x\psi(x)'-\frac{i\hbar\psi(x)}{c^\Phi_{-1}}\int_0^\infty \frac{du}{u}~\Phi(u)\Phi^*(u)'\\
&\equiv-i\hbar\frac{c^\Phi_0}{c^\Phi_{-1}}x\psi(x)'-i\hbar\lambda\psi(x) ,
\end{align}
where we have used the identities $\int_{-\infty}^{\infty}dP~P e^{i(x-y)P}=-2\pi i\,\delta'(x-y)$ and $\int_{-\infty}^{\infty}dP~e^{i(x-y)P}=2\pi\,\delta(x-y)$. Furthermore we have assumed that the fiducial vector is also square integrable with regard to the measure $x^{-2}dx$, such that $c^\Phi_0$ is finite. Note that for real fiducial vectors $\gamma$ vanishes and $\lambda$ takes the value $c^\Phi_0/2c^\Phi_{-1}$. The partial integrations that have to be performed above do not lead to boundary terms, since the fiducial vector vanishes at zero and at infinity due to $c^\Phi_{-1}$ being finite.

Further we want to compute the lower symbols for these operators,
\begin{align}
\check{R}&=\frac{ c_{0}^\Phi}{ c_{-1}^\Phi}\int_{0}^{\infty}\frac{dx}{R}~  x~   |\Phi\left( \tfrac{x}{R}\right)|^2 =\frac{c_{-3}^\Phi c_{0}^\Phi}{ c_{-1}^\Phi}\, R ,\\
\check{P}&=-i\hbar\int_{0}^{\infty}\frac{dx}{R}~ e^{-\frac{i}{\hbar}Px} \Phi^*\!\left( \tfrac{x}{R}\right)  \left( \frac{\partial}{\partial x}+\frac{\gamma}{x}\right)  e^{\frac{i}{\hbar}Px} \Phi\!\left( \tfrac{x}{R}\right)\\
&\hspace{-0.6em}\overset{v=\frac{x}{R}}{=} P - \frac{i\hbar}{R}\int_{0}^{\infty}dv~\left( |\Phi(v)|^2\right)'   = P , \\
\check{D}&=\!-i\hbar\!\int_{0}^{\infty}\!\frac{dx}{R}\, e^{-\frac{i}{\hbar}Px} \Phi^*\!\left( \tfrac{x}{R}\right) \! \left( \frac{c^\Phi_0}{c^\Phi_{-1}}x\frac{\partial}{\partial x}+\lambda \right)\!  e^{\frac{i}{\hbar}Px} \Phi\!\left( \tfrac{x}{R}\right)\\
&\hspace{-0.6em}\overset{v=\frac{x}{R}}{=} \frac{c_{-3}^\Phi c_{0}^\Phi}{ c_{-1}^\Phi}\, RP - i\hbar \lambda-i\hbar \frac{c_{0}^\Phi}{ c_{-1}^\Phi}\int_{0}^{\infty}dv \,v~\Phi^*(v)\Phi(v)'  \\
&\equiv \frac{c_{-3}^\Phi c_{0}^\Phi}{ c_{-1}^\Phi}\, D - i\hbar \alpha  ,
\end{align}
where we used the same behavior for $\Phi(x)$ at the boundaries as above to find $\check{P}$. The factor $\alpha$ is always real, and for real fiducial vectors it vanishes.


\section{Quantum corrected dynamics}\label{app:A}

Suppose our to be quantized classical theory has a Hamiltonian $H(P,R)$. The time evolution in the quantum theory is then given by the Schr\"odinger equation
\begin{equation}
i\hbar\frac{d}{d t}\ket{\psi(t)}=\hat{H}\ket{\psi(t)} .\label{eq:schroedinger}
\end{equation}
Let us further assume that one can choose $\ket{\Phi}$ such that the ACS $\ket{P,R}$ are approximately stable under the time evolution above, $\ket{P,R,t}\approx\ket{P(t),R(t)}$ with $\ket{P,R,0}=\ket{P,R}$. We want to compare the trajectories of these approximately stable ACS to the dynamics generated by $\check{H}$.

We have seen above that the parameters $R(t)$ and $P(t)$ can be extracted from $\ket{P(t),R(t)}$ by computing the lower symbols of the phase space functions $R$, $P$ and $D$. Because the momentum operator on the half line cannot be made self-adjoint, we will use here $R$ and $D$. Since the lower symbols are simply the expectation values of operators with regard to the ACS, their time evolution follows a Heisenberg equation,
\begin{align}
\dot{R}(t)&=\frac{ c_{-1}^\Phi}{c_{-3}^\Phi c_{0}^\Phi}\, \dot{\check{R}}(t)\\&=\frac{i}{\hbar}\frac{ c_{-1}^\Phi}{c_{-3}^\Phi c_{0}^\Phi}\bra{P(t),R(t)}[\hat{H},\hat{R}]\ket{P(t),R(t)} ,\label{eq:heisenbergR}\\
\dot{D}(t)&=\frac{ c_{-1}^\Phi}{c_{-3}^\Phi c_{0}^\Phi}\, \dot{\check{D}}(t)\\&=\frac{i}{\hbar}\frac{ c_{-1}^\Phi}{c_{-3}^\Phi c_{0}^\Phi}\bra{P(t),R(t)}[\hat{H},\hat{D}]\ket{P(t),R(t)} .\label{eq:heisenbergD}
\end{align}

These equations of motion can indeed be rewritten in terms of the lower symbol of the Hamiltonian $\check{H}$ as follows. First, let us write out the lower symbol of the commutator between $\hat{H}$ and $\hat{R}$, suppressing the time dependency of $R(t)$ and $P(t)$,
\begin{multline}
\bra{P,R}[\hat{H},\hat{R}]\ket{P,R}=-\frac{c^\Phi_0}{2\pi\hbar{c^\Phi_{-1}}^2}\\\int_{0}^{\infty}d\bar{R}\int_{-\infty}^\infty d\bar{P}\int_{0}^{\infty}dx\int_0^\infty dy~\frac{H(\bar{P},\bar{R})}{R\bar{R}}\\(x-y) \,e^{\frac{i}{\hbar}(x-y)(\bar{P}-P)}  \Phi\!\left(\tfrac{x}{\bar{R}}\right) \Phi^*\!\left(\tfrac{x}{R}\right) \Phi^*\!\left(\tfrac{y}{\bar{R}}\right) \Phi\!\left(\tfrac{y}{R}\right) . \label{eq:lower_symbol_commutator_R}
\end{multline}
Replacing
\begin{equation}
	(x-y) \,e^{\frac{i}{\hbar}(x-y)(\bar{P}-P)} = i\hbar \frac{\partial}{\partial P} e^{\frac{i}{\hbar}(x-y)(\bar{P}-P)}
\end{equation}
we recognize
\begin{equation}
	\bra{P,R}[\hat{H},\hat{R}]\ket{P,R}=-i\hbar\frac{c^\Phi_0}{c^\Phi_{-1}} \frac{\partial}{\partial P} \check{H}(P,R) .
\end{equation}

Analogously we proceed for $D$,
\begin{multline}
\bra{P,R}[\hat{H},\hat{D}]\ket{P,R}=-\frac{i c^\Phi_0}{2\pi{c^\Phi_{-1}}^2}\\\int_{0}^{\infty}d\bar{R}\int_{-\infty}^\infty d\bar{P}\int_{0}^{\infty}dx\int_0^\infty dy~\frac{H(\bar{P},\bar{R})}{R\bar{R}}\\ e^{-\frac{i}{\hbar}(x P + y \bar{P})}   \Phi^*\!\left(\tfrac{x}{R}\right) \Phi^*\!\left(\tfrac{y}{\bar{R}}\right) \\\left(-x\frac{\partial}{\partial x} + y\frac{\partial}{\partial y} \right)  e^{\frac{i}{\hbar}(x \bar{P} + y P)} \Phi\!\left(\tfrac{x}{\bar{R}}\right) \Phi\!\left(\tfrac{y}{R}\right) . \label{eq:lower_symbol_commutator_D}
\end{multline}
We first perform a partial integration, flipping the $x$-derivative. No boundary terms are produced since the fiducial vector vanishes at infinity and the origin. This is not an additional restriction on the fiducial vector, it follows from $c^\Phi_{-1}$ being finite. We then focus on the terms in the integral that have derivatives acting on them,
\begin{multline}
	\left(\frac{\partial}{\partial x} x + y\frac{\partial}{\partial y} \right) e^{-\frac{i}{\hbar}(x - y) P} \Phi^*\!\left(\tfrac{x}{R}\right) \Phi\!\left(\tfrac{y}{R}\right)\\=\left(1 + P\frac{\partial}{\partial P} - R \frac{\partial}{\partial R} \right)  e^{-\frac{i}{\hbar}(x - y) P} \Phi^*\!\left(\tfrac{x}{R}\right) \Phi\!\left(\tfrac{y}{R}\right) .
\end{multline}
The only step left is now to commute the derivative in $R$ with the remaining factor $1/R$ in the integral by using
\begin{equation}
\frac{1}{R}\left(1- R\frac{\partial}{\partial R}\right) f(R)  =- R\frac{\partial}{\partial R}\frac{f(R)}{R}  ,
\end{equation}
It follows for \eqref{eq:lower_symbol_commutator_D}, 
\begin{multline}
\bra{P,R}[\hat{H},\hat{D}]\ket{P,R}\\=-i\hbar\frac{c^\Phi_0}{c^\Phi_{-1}} \left(P\frac{\partial}{\partial P}- R\frac{\partial}{\partial R}\right) \check{H}(P,R) .
\end{multline}

Inserting these results into the Heisenberg equations of motion \eqref{eq:heisenbergR} and \eqref{eq:heisenbergD} gives
\begin{align}
\dot{R}(t)&=\frac{1}{c_{-3}^\Phi} \frac{\partial}{\partial P} \check{H}(P,R) .\\
\dot{D}(t)&=\frac{1}{c_{-3}^\Phi} \left(   P \frac{\partial}{\partial P} -R\frac{\partial}{\partial R} \right) \check{H}(P,R) ,
\end{align}
and replacing $D=RP$ we find
\begin{equation}
\dot{P}(t)=-\frac{1}{c_{-3}^\Phi} \frac{\partial}{\partial R} \check{H}(P,R) .\\
\end{equation}
Indeed we see that the dynamics of stable ACS are governed by the effective quantum corrected Hamiltonian $\check{H}$. For centered fiducial vectors as discussed in Sec.\ \ref{ch:chapter_3} one does not even need to rescale the lower symbol, since $c^\Phi_{-3}=1$.

It should be noted that the stability of ACS is far from a trivial problem. Since the accuracy of the semiclassical approximation as discussed here depends on this approximate stability, this topic certainly deserves further investigation. Here we do not explicitly consider this problem since we will not go beyond the quantum corrected phase space picture.


\section{Fourier transform of the squares of the hyperbolic cosecant and cotangent}\label{app:B}

Let us start with the integral
\begin{equation}
I_+(x-y)=-\int_{-\infty}^\infty d\bar{P}_A~ \frac{e^{- \frac{i}{\hbar}\bar{P}_A\left(  x - y\right) }}{\cosh^2\bar{P}_A} ,\\
\end{equation}
and solve it using contour integration. We can close the contour in the lower half plane for $x-y>0$ and in the upper half plane for $x-y<0$. The hyperbolic cosine is zero at $\bar{P}_A^0=\frac{i\pi}{2}(2k+1)$, $k\in\mathbb{Z}$, and behaves close to these roots as  $\cosh^2\bar{P}_A\sim-(\bar{P}_A-\bar{P}_A^0)^2$ such that the integrand above has second order poles there. The residue at these poles is then $-\frac{i}{\hbar} (x-y) e^{\frac{\pi}{2\hbar}(2k+1)\left(  x - y\right) }$. This gives us
\begin{align}
	I_+(x-y)&=-\frac{2\pi }{\hbar}|x-y|\sum_{k=0}^{\infty} e^{-\frac{\pi}{2\hbar}(2k+1)| x - y| }\\
	 &= -\frac{\frac{2\pi }{\hbar}|x-y| }{e^{\frac{\pi}{2\hbar}| x - y| }-e^{-\frac{\pi}{2\hbar}| x - y| }}\\
	 &= -\frac{\frac{\pi }{\hbar}(x-y) }{\sinh\left( \frac{\pi}{2\hbar}( x - y)\right)} ,
\end{align}
where we recognized the geometric series.

Analogously we proceed with the second integral
\begin{equation}
I_-(x-y)=\int_{-\infty}^\infty d\bar{P}_A~ \frac{e^{- \frac{i}{\hbar}\bar{P}_A\left(  x - y\right) }}{\sinh^2\bar{P}_A}  .\\
\end{equation}
An important difference to the first integral is that strictly speaking $I_-$ does not converge due to a singularity of the integrand at $\bar{P}_A=0$. We will regularize this integral by taking the average over two contour integrals, one where the contour on the real line is deformed to include $\bar{P}_A=0$ and one where we exclude it. The integrand has second order poles at $\bar{P}_A^0=i\pi k$, where $\sinh^2\bar{P}_A\sim(\bar{P}_A-\bar{P}_A^0)^2$, the residue is $-\frac{i}{\hbar} (x-y) e^{\frac{\pi}{\hbar}k\left(  x - y\right) }$, and we close the contours as above. The integral is then
\begin{align}
I_-(x-y)&= -\frac{2\pi }{\hbar} |x-y|\left(\frac{1}{2} + \sum_{k=1}^{\infty} e^{-\frac{\pi}{\hbar}k|x - y| } \right) \\
 &= -\frac{\pi }{\hbar} |x-y| \frac{1+e^{-\frac{\pi}{\hbar}|x - y| }}{1-e^{-\frac{\pi}{\hbar}|x - y| }} \\
 &= -\frac{\frac{\pi }{\hbar} (x-y) }{\tanh\left( \frac{\pi}{2\hbar}(x-y) \right) }  ,
\end{align}
where the first term in the sum had to be weighted with $\frac{1}{2}$ due to the regularization described above.


\section{Bouncing behavior for a different fiducial vector}\label{app:D}

To make sure our characterization of the bounce in Sec.\ \ref{ch:chapter_5} is robust we want to also check a different fiducial vector. We choose
\begin{equation}
\Phi(x)=\frac{ e^{-\frac{\ln^2 x}{4\sigma}}}{\left( 2\pi\sigma x^2\right)^\frac{1}{4}} ,
\end{equation}
where $\sigma$ is a positive real parameter. Note that the classical limit for this fiducial vector involves taking the limit $\sigma\to0$. The relevant constants are given by
\begin{equation}
c^\Phi_\alpha=e^{\frac{\sigma}{2}(2+\alpha)^2} .
\end{equation}
This fiducial vector is not centered, so we have to rescale $\check{H}_\pm$ accordingly.

For the lower symbol of the Hamiltonians \eqref{eq:hamiltonian_generic} we also need the following integral:
\begin{align}
\int_0^\infty\frac{d\bar{A}}{\sqrt{\bar{A}}} ~ &\Phi^*\left( \tfrac{x}{\bar{A}}\right) \Phi\left( \tfrac{y}{\bar{A}}\right)\\ &= \frac{1}{\sqrt{2\pi\sigma x y}} \int_0^\infty d\bar{A}~\sqrt{\bar{A}} ~e^{-\frac{1}{4\sigma} \left( \ln^2\frac{x}{A} + \ln^2\frac{y}{A}\right) }\nonumber \\
&=(x y)^\frac{1}{4}~e^{\frac{9\sigma}{8}-\frac{1}{8\sigma}\left(\ln x -\ln y \right)^2 }  .
\end{align}
This gives us
\begin{multline}
e^{-\frac{3\sigma}{4}}~\check{H}_\pm(P_A,A)=- \sqrt{\frac{A}{2}} +\frac{e^{-\frac{\sigma}{8}} }{4\hbar^2 \sqrt{\pi\sigma}}\\\int_{0}^\infty dx\int_{0}^\infty dy~  \frac{ x - y }{ F_\pm\left(\frac{\pi }{2\hbar}\left(  x - y\right) \right) } \frac{e^{\frac{i}{\hbar}P_A\left(x-y \right)}}{\left( x y\right)^\frac{1}{4}   }\\ e^{ -\frac{1}{4\sigma}\left[\frac{1}{2}\left( \ln\frac{x}{A}- \ln\frac{y}{A}\right)^2 - \ln^2\frac{x}{A}-\ln^2\frac{y}{A} \right]  }   .
\end{multline}
The function in the exponential multiplied by $1/\sigma$ has a single critical point at $x=y=A$, where it vanishes. The eigenvalues of its Hessian are $-1/A^2$ and $-1/2A^2$, meaning the critical point is a maximum. Using the saddle point approximation the Hamiltonians for $\sigma\to0$ can then be estimated as 
\begin{equation}
\check{H}_\pm(P_A,A)\sim- \sqrt{\frac{A}{2}} +  \sqrt{\frac{2\sigma A^3}{\pi\hbar^2 }}  .
\end{equation}
Under the identification $\beta=1/2\sigma$ we recover all results from the fiducial vector used in Sec.\ \ref{ch:chapter_5}.


\section{Classical limit}\label{app:C}

The classical limit in ACSQ involves not only $\hbar\to0$, but also taking a limit in your family of fiducial vectors such that $|\Phi(x)|^2\to\delta(x-1)$. For our choice of fiducial vector in Sec. \ref{ch:chapter_5} this can be achieved by $\beta\to\infty$. Taking these two limits one after the other does not produce the correct classical limit anymore, in contrast to Sec.\ \ref{ch:chapter_4}. The easiest way to obtain the correct classical limit, as also discussed in \cite{BergeronBounce}, is to let $\beta\to\infty$ as a function of $\hbar$ when taking $\hbar\to0$. Let us take a closer look at the definition of the lower symbol to find this relation: 
\begin{multline}
\check{f}=\frac{1}{2\pi \hbar c^\Phi_{-1}}\int_{0}^{\infty}dA'\int_{-\infty}^{\infty}dP_A'~f(P_A',A')\\|\braket{P_A',A'|P_A,A}|^2 .
\end{multline}
To obtain he original phase space function from the lower symbol we then need to take the limits mentioned above in such a way that
\begin{equation}
\frac{|\braket{P_A',A'|P_A,A}|^2}{2\pi \hbar c^\Phi_{-1}}\to\delta\left( A-A'\right) \,\delta\left( P_A-P_A'\right)
\end{equation}

For the fiducial vector chosen here we have
\begin{multline}
\frac{|\braket{P_A',A'|P_A,A}|^2}{2\pi \hbar c^\Phi_{-1}}=\frac{2\beta-1}{4\pi\beta\hbar}~\left[\frac{1}{4}\left(\sqrt{\frac{A'}{A}}+\sqrt{\frac{A}{A'}} \right)^2\right. \\ + \left.\vphantom{\sqrt{\frac{A'}{A}}} \frac{A\,A'}{4\beta^2\hbar^2}\left(P_A-P_A' \right)^2   \right]^{-2\beta}\label{eq:class_limit}
\end{multline}
Since the lower symbol of $1$ is again $1$, the above is normalized with regard to integration over the half plane. For $A=A'$ and $P_A=P_A'$ it reads
\begin{equation}
\frac{1}{2\pi \hbar c^\Phi_{-1}}~|\braket{P_A,A|P_A,A}|^2=\frac{2\beta-1}{4\pi\beta\hbar}
\end{equation}
and thus diverges regardless of how exactly $\beta$ depends on $\hbar$ when $\hbar\to0$. For the correct classical limit to emerge we now only need \eqref{eq:class_limit} to vanish when $A\neq A'$ or $P_A\neq P_A'$ when the limit is taken. Note to this end that
\begin{equation}
\frac{1}{4}\left(\sqrt{\frac{A'}{A}}+\sqrt{\frac{A}{A'}} \right)^2>1
\end{equation}
for all $A,A'>0$ where $A\neq A'$. Since the second term in the bracket in \eqref{eq:class_limit} is non negative, \eqref{eq:class_limit} then always vanishes exponentially for $\beta\to\infty$, as long as we do not let $\beta$ depend on $\hbar$ logarithmically. For $A= A'$ and $P_A\neq P_A'$ we only require that the product $\beta\hbar$ does not diverge. Then the term in the bracket is bigger than $1$ and the whole expression vanishes again in the limit.
The above requirements can be fulfilled by demanding that $\beta\propto1/\hbar$ for $\hbar\to0$, but other choices are also possible.


\section{Quantum corrected dynamics with modified commutation relations}\label{app:E}

Here we want to check whether our investigations of the lower symbol of the Hamiltonian from Sec.\ \ref{ch:chapter_5} carries over to the alternate quantization map leading to modified commutation relations constructed in Sec.\ \ref{ch:chapter_6}. We have already noted in Sec.\ \ref{ch:chapter_6} that one can identify operators to phase space functions $f(\Pi,R)$ in the new parametrization with operators to $f_\pm(P,R)=f(\Pi_\pm(P,R),R)$ in the old parametrization by using
\begin{equation}
\ket{P,R}_\text{old}=\ket{\Pi_\pm(P,R),R}_\text{new} .
\end{equation}
We write this identification as
\begin{equation}
\invbreve{f} = \hat{f}_\pm ,
\end{equation}
where the rounded hat denotes the quantization map using the new parametrization, and the usual hat the old one. From the above straightforwardly follows that one can also identify lower symbols,
\begin{align}
\breve{f}(\Pi_\pm(P,R),R) &= \braket{\Pi_\pm(P,R),R|\invbreve{f}|\Pi_\pm(P,R),R}_\text{new}\\ &= \braket{P,R|\hat{f}_\pm|P,R}_\text{old}  \\&= \check{f}_\pm(P,R) .
\end{align}
At the level of the lower symbols the reparametrization emerges as the original coordinate transformation on phase space.

This allows us to write,
\begin{align}
\dot{R}(t)&=\frac{1}{c_{-3}^\Phi} \frac{\partial}{\partial P} \check{H}_\pm(P,R)=\frac{1}{c_{-3}^\Phi} \frac{\partial}{\partial P} \breve{H}(\Pi_\pm(P,R),R) ,\\
\dot{P}(t)&=-\frac{1}{c_{-3}^\Phi} \frac{\partial}{\partial R} \check{H}_\pm(P,R)=-\frac{1}{c_{-3}^\Phi}  \frac{\partial}{\partial R} \breve{H}(\Pi_\pm(P,R),R) .
\end{align}
We can see that we can easily arrive at the old equations of motion from the new lower symbol of the Hamiltonian by identifying $\Pi=\Pi_\pm$, completely analogous to the classical case. Furthermore we can now see how the equations of motion look in the new phase space variables. To this end we identify
\begin{align}
\breve{P} &= \check{P}_\pm = P   ,
\end{align}
where $\breve{P}$ is given by
\begin{equation}
\breve{P} = \frac{R}{2}\ln\left|\frac{1+\frac{\Pi}{R}}{1-\frac{\Pi}{R}} \right| .
\end{equation}
We then get the equations of motion
\begin{equation}
\dot{R}(t)
= \frac{1}{c_{-3}^\Phi} \left(1-\frac{\Pi^2}{R^2} \right)  \frac{\partial}{\partial \Pi} \breve{H}(\Pi,R)   ,
\end{equation}
\begin{equation}
\dot{\Pi}(t)=
-\frac{1}{c_{-3}^\Phi} \left(1-\frac{\Pi^2}{R^2} \right) \frac{\partial}{\partial R} \breve{H}(\Pi,R) ,
\end{equation}
from a coordinate transformation $(R,P)$ to $(\Pi,R)$. The form of these equations is completely equivalent to the classical equations of motion with the modified Poisson brackets \eqref{eq:mod_bracket}.



\begin{thebibliography}{52}%
	\makeatletter
	\providecommand \@ifxundefined [1]{%
		\@ifx{#1\undefined}
	}%
	\providecommand \@ifnum [1]{%
		\ifnum #1\expandafter \@firstoftwo
		\else \expandafter \@secondoftwo
		\fi
	}%
	\providecommand \@ifx [1]{%
		\ifx #1\expandafter \@firstoftwo
		\else \expandafter \@secondoftwo
		\fi
	}%
	\providecommand \natexlab [1]{#1}%
	\providecommand \enquote  [1]{``#1''}%
	\providecommand \bibnamefont  [1]{#1}%
	\providecommand \bibfnamefont [1]{#1}%
	\providecommand \citenamefont [1]{#1}%
	\providecommand \href@noop [0]{\@secondoftwo}%
	\providecommand \href [0]{\begingroup \@sanitize@url \@href}%
	\providecommand \@href[1]{\@@startlink{#1}\@@href}%
	\providecommand \@@href[1]{\endgroup#1\@@endlink}%
	\providecommand \@sanitize@url [0]{\catcode `\\12\catcode `\$12\catcode
		`\&12\catcode `\#12\catcode `\^12\catcode `\_12\catcode `\%12\relax}%
	\providecommand \@@startlink[1]{}%
	\providecommand \@@endlink[0]{}%
	\providecommand \url  [0]{\begingroup\@sanitize@url \@url }%
	\providecommand \@url [1]{\endgroup\@href {#1}{\urlprefix }}%
	\providecommand \urlprefix  [0]{URL }%
	\providecommand \Eprint [0]{\href }%
	\providecommand \doibase [0]{http://dx.doi.org/}%
	\providecommand \selectlanguage [0]{\@gobble}%
	\providecommand \bibinfo  [0]{\@secondoftwo}%
	\providecommand \bibfield  [0]{\@secondoftwo}%
	\providecommand \translation [1]{[#1]}%
	\providecommand \BibitemOpen [0]{}%
	\providecommand \bibitemStop [0]{}%
	\providecommand \bibitemNoStop [0]{.\EOS\space}%
	\providecommand \EOS [0]{\spacefactor3000\relax}%
	\providecommand \BibitemShut  [1]{\csname bibitem#1\endcsname}%
	\let\auto@bib@innerbib\@empty
	\bibitem [{\citenamefont {Lund}(1973)}]{LundOS}%
	\BibitemOpen
	\bibfield  {author} {\bibinfo {author} {\bibfnamefont {F.}~\bibnamefont
			{Lund}},\ }\href {\doibase 10.1103/PhysRevD.8.3253} {\bibfield  {journal}
		{\bibinfo  {journal} {Phys. Rev. D}\ }\textbf {\bibinfo {volume} {8}},\
		\bibinfo {pages} {3253} (\bibinfo {year} {1973})}\BibitemShut {NoStop}%
	\bibitem [{\citenamefont {Corda}\ and\ \citenamefont {Feleppa}()}]{CordaOS}%
	\BibitemOpen
	\bibfield  {author} {\bibinfo {author} {\bibfnamefont {C.}~\bibnamefont
			{Corda}}\ and\ \bibinfo {author} {\bibfnamefont {F.}~\bibnamefont
			{Feleppa}},\ }\href@noop {} {}\Eprint {http://arxiv.org/abs/1912.06478}
	{arXiv:1912.06478 [gr-qc]} \BibitemShut {NoStop}%
	\bibitem [{\citenamefont {Peleg}(1995)}]{PelegOSspectrum}%
	\BibitemOpen
	\bibfield  {author} {\bibinfo {author} {\bibfnamefont {Y.}~\bibnamefont
			{Peleg}},\ }\href {\doibase https://doi.org/10.1016/0370-2693(95)00874-K}
	{\bibfield  {journal} {\bibinfo  {journal} {Phys. Lett. B}\ }\textbf
		{\bibinfo {volume} {356}},\ \bibinfo {pages} {462 } (\bibinfo {year}
		{1995})}\BibitemShut {NoStop}%
	\bibitem [{\citenamefont {Peleg}({\natexlab{a}})}]{PelegOS0}%
	\BibitemOpen
	\bibfield  {author} {\bibinfo {author} {\bibfnamefont {Y.}~\bibnamefont
			{Peleg}},\ }\href@noop {} {\ } \Eprint
	{http://arxiv.org/abs/9303169} {arXiv:9303169 [hep-th]} \BibitemShut
	{NoStop}%
	\bibitem [{\citenamefont {Peleg}({\natexlab{b}})}]{PelegOS}%
	\BibitemOpen
	\bibfield  {author} {\bibinfo {author} {\bibfnamefont {Y.}~\bibnamefont
			{Peleg}},\ }\href@noop {} {\ } \Eprint
	{http://arxiv.org/abs/9402036} {arXiv:9402036 [hep-th]} \BibitemShut
	{NoStop}%
	\bibitem [{\citenamefont {Schmitz}(2020)}]{MeOS}%
	\BibitemOpen
	\bibfield  {author} {\bibinfo {author} {\bibfnamefont {T.}~\bibnamefont
			{Schmitz}},\ }\href {\doibase 10.1103/PhysRevD.101.026016} {\bibfield
		{journal} {\bibinfo  {journal} {Phys. Rev. D}\ }\textbf {\bibinfo {volume}
			{101}},\ \bibinfo {pages} {026016} (\bibinfo {year} {2020})}\BibitemShut
	{NoStop}%
	\bibitem [{\citenamefont {Kiefer}\ and\ \citenamefont {Schmitz}(2019)}]{MeLTB}%
	\BibitemOpen
	\bibfield  {author} {\bibinfo {author} {\bibfnamefont {C.}~\bibnamefont
			{Kiefer}}\ and\ \bibinfo {author} {\bibfnamefont {T.}~\bibnamefont
			{Schmitz}},\ }\href {\doibase 10.1103/PhysRevD.99.126010} {\bibfield
		{journal} {\bibinfo  {journal} {Phys. Rev. D}\ }\textbf {\bibinfo {volume}
			{99}},\ \bibinfo {pages} {126010} (\bibinfo {year} {2019})}\BibitemShut
	{NoStop}%
	\bibitem [{\citenamefont {H{\'a}j{\'i}\v{c}ek}\ and\ \citenamefont
		{Kiefer}(2001)}]{HajicekKieferNullShells}%
	\BibitemOpen
	\bibfield  {author} {\bibinfo {author} {\bibfnamefont {P.}~\bibnamefont
			{H{\'a}j{\'i}\v{c}ek}}\ and\ \bibinfo {author} {\bibfnamefont
			{C.}~\bibnamefont {Kiefer}},\ }\href@noop {} {\bibfield  {journal} {\bibinfo
			{journal} {Int. J. Mod. Phys. D}\ }\textbf {\bibinfo {volume} {10}},\
		\bibinfo {pages} {775} (\bibinfo {year} {2001})}\BibitemShut {NoStop}%
	\bibitem [{\citenamefont {H{\'a}j{\'i}\v{c}ek}(2003)}]{HajicekNullShells}%
	\BibitemOpen
	\bibfield  {author} {\bibinfo {author} {\bibfnamefont {P.}~\bibnamefont
			{H{\'a}j{\'i}\v{c}ek}},\ }\enquote {\bibinfo {title} {{Quantum Theory of
				Gravitational Collapse (Lecture Notes on Quantum Conchology)}},}\ in\ \href
	{\doibase 10.1007/978-3-540-45230-0_6} {\emph {\bibinfo {booktitle} {{Quantum
					Gravity: From Theory to Experimental Search}}}},\ \bibinfo {editor} {edited
		by\ \bibinfo {editor} {\bibfnamefont {D.~J.~W.}\ \bibnamefont {Giulini}},
		\bibinfo {editor} {\bibfnamefont {C.}~\bibnamefont {Kiefer}}, \ and\ \bibinfo
		{editor} {\bibfnamefont {C.}~\bibnamefont {L{\"a}mmerzahl}}}\ (\bibinfo
	{publisher} {Springer},\ \bibinfo {address} {Berlin,
		Heidelberg},\ \bibinfo {year} {2003})\ pp.\ \bibinfo {pages}
	{255--299}\BibitemShut {NoStop}%
	\bibitem [{\citenamefont {Frolov}\ and\ \citenamefont
		{Vilkovisky}(1981)}]{FrolovNullShell}%
	\BibitemOpen
	\bibfield  {author} {\bibinfo {author} {\bibfnamefont {V.}~\bibnamefont
			{Frolov}}\ and\ \bibinfo {author} {\bibfnamefont {G.}~\bibnamefont
			{Vilkovisky}},\ }\href {\doibase
		https://doi.org/10.1016/0370-2693(81)90542-6} {\bibfield  {journal} {\bibinfo
			{journal} {Phys. Lett. B}\ }\textbf {\bibinfo {volume} {106}},\ \bibinfo
		{pages} {307 } (\bibinfo {year} {1981})}\BibitemShut {NoStop}%
	\bibitem [{\citenamefont {Ashtekar}(2009)}]{AshtekarCosmology}%
	\BibitemOpen
	\bibfield  {author} {\bibinfo {author} {\bibfnamefont {A.}~\bibnamefont
			{Ashtekar}},\ }\href {http://stacks.iop.org/1742-6596/189/i=1/a=012003}
	{\bibfield  {journal} {\bibinfo  {journal} {J. Phys.: Conf. Ser.}\
		}\textbf {\bibinfo {volume} {189}},\ \bibinfo {pages} {012003} (\bibinfo
		{year} {2009})}\BibitemShut {NoStop}%
	\bibitem [{\citenamefont {Rovelli}\ and\ \citenamefont
		{Vidotto}(2014)}]{RovelliPlanckStars}%
	\BibitemOpen
	\bibfield  {author} {\bibinfo {author} {\bibfnamefont {C.}~\bibnamefont
			{Rovelli}}\ and\ \bibinfo {author} {\bibfnamefont {F.}~\bibnamefont
			{Vidotto}},\ }\href {\doibase 10.1142/S0218271814420267} {\bibfield
		{journal} {\bibinfo  {journal} {Int. J. Mod. Phys. D}\ }\textbf {\bibinfo
			{volume} {23}},\ \bibinfo {pages} {1442026} (\bibinfo {year}
		{2014})}\BibitemShut {NoStop}%
	\bibitem [{\citenamefont {Bojowald}(2020)}]{BojowaldCritique}%
	\BibitemOpen
	\bibfield  {author} {\bibinfo {author} {\bibfnamefont {M.}~\bibnamefont
		{Bojowald}},\ }\href {\doibase 10.3390/universe6030036} {\bibfield
	{journal} {\bibinfo  {journal} {Universe}\ }\textbf {\bibinfo {volume} {6}},\
	\bibinfo {pages} {36} (\bibinfo {year} {2020})}\BibitemShut {NoStop}%
	\bibitem [{\citenamefont {Bambi}\ \emph {et~al.}(2013)\citenamefont {Bambi},
		\citenamefont {Malafarina},\ and\ \citenamefont
		{Modesto}}]{MalafarinaBounce}%
	\BibitemOpen
	\bibfield  {author} {\bibinfo {author} {\bibfnamefont {C.}~\bibnamefont
			{Bambi}}, \bibinfo {author} {\bibfnamefont {D.}~\bibnamefont {Malafarina}}, \
		and\ \bibinfo {author} {\bibfnamefont {L.}~\bibnamefont {Modesto}},\ }\href
	{\doibase 10.1103/PhysRevD.88.044009} {\bibfield  {journal} {\bibinfo
			{journal} {Phys. Rev. D}\ }\textbf {\bibinfo {volume} {88}},\ \bibinfo
		{pages} {044009} (\bibinfo {year} {2013})}\BibitemShut {NoStop}%
	\bibitem [{\citenamefont {Liu}\ \emph {et~al.}(2014)\citenamefont {Liu},
		\citenamefont {Malafarina}, \citenamefont {Modesto},\ and\ \citenamefont
		{Bambi}}]{LiuBounce}%
	\BibitemOpen
	\bibfield  {author} {\bibinfo {author} {\bibfnamefont {Y.}~\bibnamefont
			{Liu}}, \bibinfo {author} {\bibfnamefont {D.}~\bibnamefont {Malafarina}},
		\bibinfo {author} {\bibfnamefont {L.}~\bibnamefont {Modesto}}, \ and\
		\bibinfo {author} {\bibfnamefont {C.}~\bibnamefont {Bambi}},\ }\href
	{\doibase 10.1103/PhysRevD.90.044040} {\bibfield  {journal} {\bibinfo
			{journal} {Phys. Rev. D}\ }\textbf {\bibinfo {volume} {90}},\ \bibinfo
		{pages} {044040} (\bibinfo {year} {2014})}\BibitemShut {NoStop}%
	\bibitem [{\citenamefont {{Ben Achour}}\ \emph {et~al.}()\citenamefont {{Ben
				Achour}}, \citenamefont {Brahma},\ and\ \citenamefont
		{Uzan}}]{AchourBounce1}%
	\BibitemOpen
	\bibfield  {author} {\bibinfo {author} {\bibfnamefont {J.}~\bibnamefont {{Ben
					Achour}}}, \bibinfo {author} {\bibfnamefont {S.}~\bibnamefont {Brahma}}, \
		and\ \bibinfo {author} {\bibfnamefont {J.-P.}\ \bibnamefont {Uzan}},\
	}\href {\doibase 10.1088/1475-7516/2020/03/041} {\bibfield  {journal} {\bibinfo
	{journal} {J. Cosmol. Astropart. Phys.}\ }\textbf {\bibinfo {volume} {03}},\ \bibinfo
	{pages} {041} (\bibinfo {year} {2020})}\BibitemShut {NoStop}%
	\bibitem [{\citenamefont {{Ben Achour}}\ and\ \citenamefont
		{Uzan}()}]{AchourBounce2}%
	\BibitemOpen
	\bibfield  {author} {\bibinfo {author} {\bibfnamefont {J.}~\bibnamefont {{Ben
					Achour}}}\ and\ \bibinfo {author} {\bibfnamefont {J.-P.}\ \bibnamefont
			{Uzan}},\ }\href@noop {} {}\Eprint {http://arxiv.org/abs/2001.06153}
	{arXiv:2001.06153 [gr-qc]} \BibitemShut {NoStop}%
		\bibitem [{\citenamefont {{Ben Achour}}\ \emph {et~al.}()\citenamefont {{Ben
				Achour}}, \citenamefont {Brahma}, \citenamefont {Mukohyama},\ and\
		\citenamefont {Uzan}}]{AchourBounce3}%
	\BibitemOpen
	\bibfield  {author} {\bibinfo {author} {\bibfnamefont {J.}~\bibnamefont {{Ben
					Achour}}}, \bibinfo {author} {\bibfnamefont {S.}~\bibnamefont {Brahma}},
		\bibinfo {author} {\bibfnamefont {S.}~\bibnamefont {Mukohyama}}, \ and\
		\bibinfo {author} {\bibfnamefont {J.-P.}\ \bibnamefont {Uzan}},\ }\href@noop
	{} {}\Eprint {http://arxiv.org/abs/2004.12977} {arXiv:2004.12977 [gr-qc]}
	\BibitemShut {NoStop}%
	\bibitem [{\citenamefont {Ambrus}\ and\ \citenamefont
		{H{\'a}j{\'i}\v{c}ek}(2005)}]{AmbrusHajicekLifetime}%
	\BibitemOpen
	\bibfield  {author} {\bibinfo {author} {\bibfnamefont {M.}~\bibnamefont
			{Ambrus}}\ and\ \bibinfo {author} {\bibfnamefont {P.}~\bibnamefont
			{H{\'a}j{\'i}\v{c}ek}},\ }\href {\doibase 10.1103/PhysRevD.72.064025}
	{\bibfield  {journal} {\bibinfo  {journal} {Phys. Rev. D}\ }\textbf {\bibinfo
			{volume} {72}},\ \bibinfo {pages} {064025} (\bibinfo {year}
		{2005})}\BibitemShut {NoStop}%
	\bibitem [{\citenamefont {Christodoulou}\ and\ \citenamefont
		{D'Ambrosio}()}]{ChristodoulouLifetime}%
	\BibitemOpen
	\bibfield  {author} {\bibinfo {author} {\bibfnamefont {M.}~\bibnamefont
			{Christodoulou}}\ and\ \bibinfo {author} {\bibfnamefont {F.}~\bibnamefont
			{D'Ambrosio}},\ }\href@noop {} {}\Eprint {http://arxiv.org/abs/1801.03027}
	{arXiv:1801.03027 [gr-qc]} \BibitemShut {NoStop}%
	\bibitem [{\citenamefont {Christodoulou}\ \emph {et~al.}(2016)\citenamefont
		{Christodoulou}, \citenamefont {Rovelli}, \citenamefont {Speziale},\ and\
		\citenamefont {Vilensky}}]{ChristodoulouLifetime2}%
	\BibitemOpen
	\bibfield  {author} {\bibinfo {author} {\bibfnamefont {M.}~\bibnamefont
			{Christodoulou}}, \bibinfo {author} {\bibfnamefont {C.}~\bibnamefont
			{Rovelli}}, \bibinfo {author} {\bibfnamefont {S.}~\bibnamefont {Speziale}}, \
		and\ \bibinfo {author} {\bibfnamefont {I.}~\bibnamefont {Vilensky}},\ }\href
	{\doibase 10.1103/PhysRevD.94.084035} {\bibfield  {journal} {\bibinfo
			{journal} {Phys. Rev. D}\ }\textbf {\bibinfo {volume} {94}},\ \bibinfo
		{pages} {084035} (\bibinfo {year} {2016})}\BibitemShut {NoStop}%
	\bibitem [{\citenamefont {Barcel{\'o}}\ \emph {et~al.}(2017)\citenamefont
		{Barcel{\'o}}, \citenamefont {Carballo-Rubio},\ and\ \citenamefont
		{Garay}}]{BarceloLifetime}%
	\BibitemOpen
	\bibfield  {author} {\bibinfo {author} {\bibfnamefont {C.}~\bibnamefont
			{Barcel{\'o}}}, \bibinfo {author} {\bibfnamefont {R.}~\bibnamefont
			{Carballo-Rubio}}, \ and\ \bibinfo {author} {\bibfnamefont {L.~J.}\
			\bibnamefont {Garay}},\ }\href
	{http://stacks.iop.org/0264-9381/34/i=10/a=105007} {\bibfield  {journal}
		{\bibinfo  {journal} {Class. Quantum Grav.}\ }\textbf {\bibinfo {volume}
			{34}},\ \bibinfo {pages} {105007} (\bibinfo {year} {2017})}\BibitemShut
	{NoStop}%
	\bibitem [{\citenamefont {Barcel{\'o}}\ \emph {et~al.}(2016)\citenamefont
		{Barcel{\'o}}, \citenamefont {Carballo-Rubio},\ and\ \citenamefont
		{Garay}}]{BarceloBounce1}%
	\BibitemOpen
	\bibfield  {author} {\bibinfo {author} {\bibfnamefont {C.}~\bibnamefont
			{Barcel{\'o}}}, \bibinfo {author} {\bibfnamefont {R.}~\bibnamefont
			{Carballo-Rubio}}, \ and\ \bibinfo {author} {\bibfnamefont {L.~J.}\
			\bibnamefont {Garay}},\ }\href {\doibase 10.1007/JHEP01(2016)157} {\bibfield
		{journal} {\bibinfo  {journal} {J. High Energy Phys.}\ }\textbf {\bibinfo
			{volume} {01}},\ \bibinfo {pages} {157} (\bibinfo {year} {2016})}\BibitemShut
	{NoStop}%
	\bibitem [{\citenamefont {Bambi}\ \emph {et~al.}(2014)\citenamefont {Bambi},
		\citenamefont {Malafarina},\ and\ \citenamefont {Modesto}}]{BambiBounce}%
	\BibitemOpen
	\bibfield  {author} {\bibinfo {author} {\bibfnamefont {C.}~\bibnamefont
			{Bambi}}, \bibinfo {author} {\bibfnamefont {D.}~\bibnamefont {Malafarina}}, \
		and\ \bibinfo {author} {\bibfnamefont {L.}~\bibnamefont {Modesto}},\ }\href
	{\doibase 10.1140/epjc/s10052-014-2767-9} {\bibfield  {journal} {\bibinfo
			{journal} {Eur. Phys. J. C}\ }\textbf {\bibinfo {volume} {74}},\ \bibinfo
		{pages} {2767} (\bibinfo {year} {2014})}\BibitemShut {NoStop}%
	\bibitem [{\citenamefont
		{H{\'a}j{\'i}\v{c}ek}(2001)}]{HajicekQuantumNullShells}%
	\BibitemOpen
	\bibfield  {author} {\bibinfo {author} {\bibfnamefont {P.}~\bibnamefont
			{H{\'a}j{\'i}\v{c}ek}},\ }\href {\doibase 10.1016/S0550-3213(01)00140-7}
	{\bibfield  {journal} {\bibinfo  {journal} {Nucl. Phys. B}\ }\textbf
		{\bibinfo {volume} {603}},\ \bibinfo {pages} {555} (\bibinfo {year}
		{2001})}\BibitemShut {NoStop}%
	\bibitem [{\citenamefont {Barcel{\'o}}\ \emph {et~al.}(2015)\citenamefont
		{Barcel{\'o}}, \citenamefont {Carballo-Rubio}, \citenamefont {Garay},\ and\
		\citenamefont {Jannes}}]{BarceloBounce2}%
	\BibitemOpen
	\bibfield  {author} {\bibinfo {author} {\bibfnamefont {C.}~\bibnamefont
			{Barcel{\'o}}}, \bibinfo {author} {\bibfnamefont {R.}~\bibnamefont
			{Carballo-Rubio}}, \bibinfo {author} {\bibfnamefont {L.~J.}\ \bibnamefont
			{Garay}}, \ and\ \bibinfo {author} {\bibfnamefont {G.}~\bibnamefont
			{Jannes}},\ }\href {http://stacks.iop.org/0264-9381/32/i=3/a=035012}
	{\bibfield  {journal} {\bibinfo  {journal} {Class. Quantum Grav.}\ }\textbf
		{\bibinfo {volume} {32}},\ \bibinfo {pages} {035012} (\bibinfo {year}
		{2015})}\BibitemShut {NoStop}%
	\bibitem [{\citenamefont {Barcel{\'o}}\ \emph {et~al.}(2014)\citenamefont
		{Barcel{\'o}}, \citenamefont {Carballo-Rubio},\ and\ \citenamefont
		{Garay}}]{BarceloBounce3}%
	\BibitemOpen
	\bibfield  {author} {\bibinfo {author} {\bibfnamefont {C.}~\bibnamefont
			{Barcel{\'o}}}, \bibinfo {author} {\bibfnamefont {R.}~\bibnamefont
			{Carballo-Rubio}}, \ and\ \bibinfo {author} {\bibfnamefont {L.~J.}\
			\bibnamefont {Garay}},\ }\href {\doibase 10.1142/S021827181442022X}
	{\bibfield  {journal} {\bibinfo  {journal} {Int. J. Mod. Phys. D}\ }\textbf
		{\bibinfo {volume} {23}},\ \bibinfo {pages} {1442022} (\bibinfo {year}
		{2014})}\BibitemShut {NoStop}%
	\bibitem [{\citenamefont {Haggard}\ and\ \citenamefont
		{Rovelli}(2015)}]{HaggardRovelliBounce}%
	\BibitemOpen
	\bibfield  {author} {\bibinfo {author} {\bibfnamefont {H.~M.}\ \bibnamefont
			{Haggard}}\ and\ \bibinfo {author} {\bibfnamefont {C.}~\bibnamefont
			{Rovelli}},\ }\href {\doibase 10.1103/PhysRevD.92.104020} {\bibfield
		{journal} {\bibinfo  {journal} {Phys. Rev. D}\ }\textbf {\bibinfo {volume}
			{92}},\ \bibinfo {pages} {104020} (\bibinfo {year} {2015})}\BibitemShut
	{NoStop}%
	\bibitem [{\citenamefont {Malafarina}(2017)}]{MalafarinaBounceRev}%
	\BibitemOpen
	\bibfield  {author} {\bibinfo {author} {\bibfnamefont {D.}~\bibnamefont
			{Malafarina}},\ }\href {\doibase 10.3390/universe3020048} {\bibfield
		{journal} {\bibinfo  {journal} {Universe}\ }\textbf {\bibinfo {volume} {3}},\
		\bibinfo {pages} {48} (\bibinfo {year} {2017})}\BibitemShut {NoStop}%
	\bibitem [{\citenamefont {Bergeron}\ and\ \citenamefont
		{Gazeau}(2014)}]{BergeronCSQ}%
	\BibitemOpen
	\bibfield  {author} {\bibinfo {author} {\bibfnamefont {H.}~\bibnamefont
			{Bergeron}}\ and\ \bibinfo {author} {\bibfnamefont {J.}~\bibnamefont
			{Gazeau}},\ }\href {\doibase https://doi.org/10.1016/j.aop.2014.02.008}
	{\bibfield  {journal} {\bibinfo  {journal} {Ann. Phys. (NY)}\ }\textbf {\bibinfo
			{volume} {344}},\ \bibinfo {pages} {43 } (\bibinfo {year}
		{2014})}\BibitemShut {NoStop}%
	\bibitem [{\citenamefont {Gazeau}\ and\ \citenamefont
		{Murenzi}(2016)}]{GazeauACSQ}%
	\BibitemOpen
	\bibfield  {author} {\bibinfo {author} {\bibfnamefont {J.~P.}\ \bibnamefont
			{Gazeau}}\ and\ \bibinfo {author} {\bibfnamefont {R.}~\bibnamefont
			{Murenzi}},\ }\href {\doibase 10.1063/1.4949366} {\bibfield  {journal}
		{\bibinfo  {journal} {J. Math. Phys.}\ }\textbf {\bibinfo {volume} {57}},\
		\bibinfo {pages} {052102} (\bibinfo {year} {2016})}\BibitemShut {NoStop}%
	\bibitem [{\citenamefont {Almeida}\ \emph {et~al.}(2018)\citenamefont
		{Almeida}, \citenamefont {Bergeron}, \citenamefont {Gazeau},\ and\
		\citenamefont {Scardua}}]{AlmeidaACS}%
	\BibitemOpen
	\bibfield  {author} {\bibinfo {author} {\bibfnamefont {C.~R.}\ \bibnamefont
			{Almeida}}, \bibinfo {author} {\bibfnamefont {H.}~\bibnamefont {Bergeron}},
		\bibinfo {author} {\bibfnamefont {J.~P.}\ \bibnamefont {Gazeau}}, \ and\
		\bibinfo {author} {\bibfnamefont {A.~C.}\ \bibnamefont {Scardua}},\ }\href
	{\doibase https://doi.org/10.1016/j.aop.2018.03.010} {\bibfield  {journal}
		{\bibinfo  {journal} {Ann. Phys. (NY)}\ }\textbf {\bibinfo {volume} {392}},\
		\bibinfo {pages} {206 } (\bibinfo {year} {2018})}\BibitemShut {NoStop}%
	\bibitem [{\citenamefont {Bergeron}\ \emph {et~al.}(2014)\citenamefont
		{Bergeron}, \citenamefont {Dapor}, \citenamefont {Gazeau},\ and\
		\citenamefont {Ma\l{}kiewicz}}]{BergeronBounce}%
	\BibitemOpen
	\bibfield  {author} {\bibinfo {author} {\bibfnamefont {H.}~\bibnamefont
			{Bergeron}}, \bibinfo {author} {\bibfnamefont {A.}~\bibnamefont {Dapor}},
		\bibinfo {author} {\bibfnamefont {J.~P.}\ \bibnamefont {Gazeau}}, \ and\
		\bibinfo {author} {\bibfnamefont {P.}~\bibnamefont {Ma\l{}kiewicz}},\ }\href
	{\doibase 10.1103/PhysRevD.89.083522} {\bibfield  {journal} {\bibinfo
			{journal} {Phys. Rev. D}\ }\textbf {\bibinfo {volume} {89}},\ \bibinfo
		{pages} {083522} (\bibinfo {year} {2014})}\BibitemShut {NoStop}%
	\bibitem [{\citenamefont {Bergeron}\ \emph {et~al.}(2019)\citenamefont
		{Bergeron}, \citenamefont {Czuchry}, \citenamefont {Gazeau},\ and\
		\citenamefont {Ma{\l}kiewicz}}]{BergeronMixmaster}%
	\BibitemOpen
	\bibfield  {author} {\bibinfo {author} {\bibfnamefont {H.}~\bibnamefont
			{Bergeron}}, \bibinfo {author} {\bibfnamefont {E.}~\bibnamefont {Czuchry}},
		\bibinfo {author} {\bibfnamefont {J.~P.}\ \bibnamefont {Gazeau}}, \ and\
		\bibinfo {author} {\bibfnamefont {P.}~\bibnamefont {Ma{\l}kiewicz}},\
	}\href@noop {} {\bibfield  {journal} {\bibinfo  {journal} {Universe}\
		}\textbf {\bibinfo {volume} {6}},\ \bibinfo {pages} {7} (\bibinfo {year}
		{2019})}\BibitemShut {NoStop}%
	\bibitem [{\citenamefont {G{\'o}{\'{z}}d{\'{z}}}\ \emph
		{et~al.}(2019)\citenamefont {G{\'o}{\'{z}}d{\'{z}}}, \citenamefont
		{Piechocki},\ and\ \citenamefont {Plewa}}]{GozdzBKL}%
	\BibitemOpen
	\bibfield  {author} {\bibinfo {author} {\bibfnamefont {A.}~\bibnamefont
			{G{\'o}{\'{z}}d{\'{z}}}}, \bibinfo {author} {\bibfnamefont {W.}~\bibnamefont
			{Piechocki}}, \ and\ \bibinfo {author} {\bibfnamefont {G.}~\bibnamefont
			{Plewa}},\ }\href {\doibase 10.1140/epjc/s10052-019-6571-4} {\bibfield
		{journal} {\bibinfo  {journal} {Eur. Phys. J. C}\ }\textbf {\bibinfo {volume}
			{79}},\ \bibinfo {pages} {45} (\bibinfo {year} {2019})}\BibitemShut {NoStop}%
	\bibitem [{\citenamefont {G{\'o}{\'z}d{\'z}}\ and\ \citenamefont
		{Piechocki}(2020)}]{GozdzBKL2}%
	\BibitemOpen
	\bibfield  {author} {\bibinfo {author} {\bibfnamefont {A.}~\bibnamefont
			{G{\'o}{\'z}d{\'z}}}\ and\ \bibinfo {author} {\bibfnamefont {W.}~\bibnamefont
			{Piechocki}},\ }\href {\doibase 10.1140/epjc/s10052-020-7668-5} {\bibfield
		{journal} {\bibinfo  {journal} {Eur. Phys. J. C}\ }\textbf {\bibinfo {volume}
			{80}},\ \bibinfo {pages} {142} (\bibinfo {year} {2020})}\BibitemShut
	{NoStop}%
	\bibitem [{\citenamefont {Kucha\v{r}}(1994)}]{KucharSchwarzschild}%
	\BibitemOpen
	\bibfield  {author} {\bibinfo {author} {\bibfnamefont {K.~V.}\ \bibnamefont
			{Kucha\v{r}}},\ }\href {\doibase 10.1103/PhysRevD.50.3961} {\bibfield
		{journal} {\bibinfo  {journal} {Phys. Rev. D}\ }\textbf {\bibinfo {volume}
			{50}},\ \bibinfo {pages} {3961} (\bibinfo {year} {1994})}\BibitemShut
	{NoStop}%
	\bibitem [{\citenamefont {Brown}\ and\ \citenamefont
		{Kucha\v{r}}(1995)}]{KucharBrownDust}%
	\BibitemOpen
	\bibfield  {author} {\bibinfo {author} {\bibfnamefont {J.~D.}\ \bibnamefont
			{Brown}}\ and\ \bibinfo {author} {\bibfnamefont {K.~V.}\ \bibnamefont
			{Kucha\v{r}}},\ }\href {\doibase 10.1103/PhysRevD.51.5600} {\bibfield
		{journal} {\bibinfo  {journal} {Phys. Rev. D}\ }\textbf {\bibinfo {volume}
			{51}},\ \bibinfo {pages} {5600} (\bibinfo {year} {1995})}\BibitemShut
	{NoStop}%
	\bibitem [{\citenamefont {H\'aj\'i\v{c}ek}\ and\ \citenamefont
		{Kijowski}(1998)}]{HajicekKijowskiFluid}%
	\BibitemOpen
	\bibfield  {author} {\bibinfo {author} {\bibfnamefont {P.}~\bibnamefont
			{H\'aj\'i\v{c}ek}}\ and\ \bibinfo {author} {\bibfnamefont {J.}~\bibnamefont
			{Kijowski}},\ }\href {\doibase 10.1103/PhysRevD.57.914} {\bibfield  {journal}
		{\bibinfo  {journal} {Phys. Rev. D}\ }\textbf {\bibinfo {volume} {57}},\
		\bibinfo {pages} {914} (\bibinfo {year} {1998})}\BibitemShut {NoStop}%
	\bibitem [{\citenamefont {Kwidzinski}\ \emph {et~al.}(2020)\citenamefont {Kwidzinski},
		\citenamefont {Malafarina},\citenamefont {Ostrowski},
		\citenamefont {Piechocki},\ and\ \citenamefont
		{Schmitz}}]{MeHamiltonian}%
	\BibitemOpen
	\bibfield  {author} {\bibinfo {author} {\bibfnamefont {N.}~\bibnamefont
			{Kwidzinski}}, \bibinfo {author} {\bibfnamefont {D.}~\bibnamefont {Malafarina}}, \bibinfo {author} {\bibfnamefont {J.~J.}~\bibnamefont {Ostrowski}}, \bibinfo {author} {\bibfnamefont {W.}~\bibnamefont {Piechocki}}, \
		and\ \bibinfo {author} {\bibfnamefont {T.}~\bibnamefont {Schmitz}},\ }\href
	{\doibase 10.1103/PhysRevD.101.104017} {\bibfield  {journal} {\bibinfo
			{journal} {Phys. Rev. D}\ }\textbf {\bibinfo {volume} {101}},\ \bibinfo
		{pages} {104017} (\bibinfo {year} {2020})}\BibitemShut {NoStop}%
	\bibitem [{\citenamefont {Martel}\ and\ \citenamefont
		{Poisson}(2001)}]{MartelCoordinates}%
	\BibitemOpen
	\bibfield  {author} {\bibinfo {author} {\bibfnamefont {K.}~\bibnamefont
			{Martel}}\ and\ \bibinfo {author} {\bibfnamefont {E.}~\bibnamefont
			{Poisson}},\ }\href@noop {} {\bibfield  {journal} {\bibinfo  {journal} {Am.
				J. Phys.}\ }\textbf {\bibinfo {volume} {69}},\ \bibinfo {pages} {476}
		(\bibinfo {year} {2001})}\BibitemShut {NoStop}%
	\bibitem [{\citenamefont {Gautreau}\ and\ \citenamefont
		{Hoffmann}(1978)}]{GautreauCoordinates}%
	\BibitemOpen
	\bibfield  {author} {\bibinfo {author} {\bibfnamefont {R.}~\bibnamefont
			{Gautreau}}\ and\ \bibinfo {author} {\bibfnamefont {B.}~\bibnamefont
			{Hoffmann}},\ }\href@noop {} {\bibfield  {journal} {\bibinfo  {journal}
			{Phys. Rev. D}\ }\textbf {\bibinfo {volume} {17}},\ \bibinfo {pages} {2552}
		(\bibinfo {year} {1978})}\BibitemShut {NoStop}%
	\bibitem [{\citenamefont {Perelomov}(1972)}]{PerelomovLieCS}%
	\BibitemOpen
	\bibfield  {author} {\bibinfo {author} {\bibfnamefont {A.~M.}\ \bibnamefont
			{Perelomov}},\ }\href {\doibase 10.1007/BF01645091} {\bibfield  {journal}
		{\bibinfo  {journal} {Commun. Math. Phys.}\ }\textbf {\bibinfo {volume}
			{26}},\ \bibinfo {pages} {222} (\bibinfo {year} {1972})}\BibitemShut
	{NoStop}%
	\bibitem [{\citenamefont {G\'{o}\'{z}d\'{z}}\ \emph {et~al.}()\citenamefont
		{G\'{o}\'{z}d\'{z}}, \citenamefont {Piechocki},\ and\ \citenamefont
		{Schmitz}}]{MeParameterizations}%
	\BibitemOpen
	\bibfield  {author} {\bibinfo {author} {\bibfnamefont {A.}~\bibnamefont
			{G\'{o}\'{z}d\'{z}}}, \bibinfo {author} {\bibfnamefont {W.}~\bibnamefont
			{Piechocki}}, \ and\ \bibinfo {author} {\bibfnamefont {T.}~\bibnamefont
			{Schmitz}},\ }\href@noop {} {}\Eprint {http://arxiv.org/abs/1908.10039}
	{arXiv:1908.10039 [math-ph]} \BibitemShut {NoStop}%
	\bibitem [{\citenamefont {Klauder}(2015)}]{KlauderEnhanced}%
	\BibitemOpen
	\bibfield  {author} {\bibinfo {author} {\bibfnamefont {J.~R.}\ \bibnamefont
			{Klauder}},\ }\href {\doibase 10.1142/9452} {\emph {\bibinfo {title}
			{Enhanced Quantization}}}\ (\bibinfo  {publisher} {World Scientific, Singapore},\
	\bibinfo {year} {2015})\BibitemShut {NoStop}%
	\bibitem [{\citenamefont {Weissman}(1983)}]{WeissmanCS}%
	\BibitemOpen
	\bibfield  {author} {\bibinfo {author} {\bibfnamefont {Y.}~\bibnamefont
			{Weissman}},\ }\href {\doibase 10.1088/0305-4470/16/12/016} {\bibfield
		{journal} {\bibinfo  {journal} {J. Phys. A: Math. Gen.}\ }\textbf {\bibinfo
			{volume} {16}},\ \bibinfo {pages} {2693} (\bibinfo {year}
		{1983})}\BibitemShut {NoStop}%
	\bibitem [{\citenamefont {Klauder}(1979)}]{KlauderCS}%
	\BibitemOpen
	\bibfield  {author} {\bibinfo {author} {\bibfnamefont {J.~R.}\ \bibnamefont
			{Klauder}},\ }\href {\doibase 10.1103/PhysRevD.19.2349} {\bibfield  {journal}
		{\bibinfo  {journal} {Phys. Rev. D}\ }\textbf {\bibinfo {volume} {19}},\
		\bibinfo {pages} {2349} (\bibinfo {year} {1979})}\BibitemShut {NoStop}%
		\bibitem [{\citenamefont {Casadio}(2000)}]{CasadioOSBounce}%
	\BibitemOpen
	\bibfield  {author} {\bibinfo {author} {\bibfnamefont {R.}~\bibnamefont
			{Casadio}},\ }\href@noop {} {\bibfield  {journal} {\bibinfo
			{journal} {Int. J. Mod. Phys. D}\ }\textbf {\bibinfo {volume} {09}},\
		\bibinfo {pages} {511} (\bibinfo {year} {2000})}\BibitemShut {NoStop}%
	\bibitem [{\citenamefont {Shapere}\ and\ \citenamefont
		{Wilczek}(2012)}]{ShapereMultivalued}%
	\BibitemOpen
	\bibfield  {author} {\bibinfo {author} {\bibfnamefont {A.}~\bibnamefont
			{Shapere}}\ and\ \bibinfo {author} {\bibfnamefont {F.}~\bibnamefont
			{Wilczek}},\ }\href {\doibase 10.1103/PhysRevLett.109.200402} {\bibfield
		{journal} {\bibinfo  {journal} {Phys. Rev. Lett.}\ }\textbf {\bibinfo
			{volume} {109}},\ \bibinfo {pages} {200402} (\bibinfo {year}
		{2012})}\BibitemShut {NoStop}%
	\bibitem [{\citenamefont {Zhao}\ \emph {et~al.}(2013)\citenamefont {Zhao},
		\citenamefont {Yu},\ and\ \citenamefont {Xu}}]{LiuMultivalued}%
	\BibitemOpen
	\bibfield  {author} {\bibinfo {author} {\bibfnamefont {L.}~\bibnamefont
			{Zhao}}, \bibinfo {author} {\bibfnamefont {P.}~\bibnamefont {Yu}}, \ and\
		\bibinfo {author} {\bibfnamefont {W.}~\bibnamefont {Xu}},\ }\href {\doibase
		10.1142/S0217732313500028} {\bibfield  {journal} {\bibinfo  {journal} {Mod.
				Phys. Lett. A}\ }\textbf {\bibinfo {volume} {28}},\ \bibinfo {pages}
		{1350002} (\bibinfo {year} {2013})}\BibitemShut {NoStop}%
	\bibitem [{\citenamefont {Ruz}\ \emph {et~al.}(2016)\citenamefont {Ruz},
		\citenamefont {Mandal}, \citenamefont {Debnath},\ and\ \citenamefont
		{Sanyal}}]{RuzMultivalued}%
	\BibitemOpen
	\bibfield  {author} {\bibinfo {author} {\bibfnamefont {S.}~\bibnamefont
			{Ruz}}, \bibinfo {author} {\bibfnamefont {R.}~\bibnamefont {Mandal}},
		\bibinfo {author} {\bibfnamefont {S.}~\bibnamefont {Debnath}}, \ and\
		\bibinfo {author} {\bibfnamefont {A.~K.}\ \bibnamefont {Sanyal}},\ }\href
	{\doibase 10.1007/s10714-016-2080-z} {\bibfield  {journal} {\bibinfo
			{journal} {Gen. Relativ. Gravit.}\ }\textbf {\bibinfo {volume} {48}},\
		\bibinfo {pages} {86} (\bibinfo {year} {2016})}\BibitemShut {NoStop}%
	\bibitem [{\citenamefont {Henneaux}\ \emph {et~al.}(1987)\citenamefont
		{Henneaux}, \citenamefont {Teitelboim},\ and\ \citenamefont
		{Zanelli}}]{HenneauxMultivalued}%
	\BibitemOpen
	\bibfield  {author} {\bibinfo {author} {\bibfnamefont {M.}~\bibnamefont
			{Henneaux}}, \bibinfo {author} {\bibfnamefont {C.}~\bibnamefont
			{Teitelboim}}, \ and\ \bibinfo {author} {\bibfnamefont {J.}~\bibnamefont
			{Zanelli}},\ }\href {\doibase 10.1103/PhysRevA.36.4417} {\bibfield  {journal}
		{\bibinfo  {journal} {Phys. Rev. A}\ }\textbf {\bibinfo {volume} {36}},\
		\bibinfo {pages} {4417} (\bibinfo {year} {1987})}\BibitemShut {NoStop}%
	\bibitem [{{\relax DLMF}()}]{DLMF}%
	\BibitemOpen
	{\relax DLMF},\ \href {http://dlmf.nist.gov/} {\enquote {\bibinfo {title}
			{\textit{NIST Digital Library of Mathematical Functions}},}\ }\bibinfo
	{howpublished} {http://dlmf.nist.gov/, Release 1.0.25 of 2019-12-15},\
	\bibinfo {note} {f.~W.~J. Olver, A.~B. {Olde Daalhuis}, D.~W. Lozier, B.~I.
		Schneider, R.~F. Boisvert, C.~W. Clark, B.~R. Miller, B.~V. Saunders, H.~S.
		Cohl, and M.~A. McClain, eds.}\BibitemShut {Stop}%
	\bibitem [{\citenamefont {Mignemi}(2012)}]{MignemiSnyderSpace}%
	\BibitemOpen
	\bibfield  {author} {\bibinfo {author} {\bibfnamefont {S.}~\bibnamefont
			{Mignemi}},\ }\href {\doibase 10.1088/1742-6596/343/1/012074} {\bibfield
		{journal} {\bibinfo  {journal} {J. Phys.: Conf. Ser.}\ }\textbf
		{\bibinfo {volume} {343}},\ \bibinfo {pages} {012074} (\bibinfo {year}
		{2012})}\BibitemShut {NoStop}%
	\bibitem [{\citenamefont {Klauder}(1994)}]{KlauderCanonical}%
	\BibitemOpen
	\bibfield  {author} {\bibinfo {author} {\bibfnamefont {J.~R.}\ \bibnamefont
			{Klauder}},\ }\href@noop {} {\bibfield  {journal} {\bibinfo  {journal} {Int.
				J. Theor. Phys.}\ }\textbf {\bibinfo {volume} {33}},\ \bibinfo {pages} {509}
		(\bibinfo {year} {1994})}\BibitemShut {NoStop}%
	\bibitem [{\citenamefont {Vanrietvelde}\ \emph {et~al.}(2020)\citenamefont
		{Vanrietvelde}, \citenamefont {Hoehn}, \citenamefont {Giacomini},\ and\
		\citenamefont {Castro-Ruiz}}]{VanrietveldeRelational}%
	\BibitemOpen
	\bibfield  {author} {\bibinfo {author} {\bibfnamefont {A.}~\bibnamefont
			{Vanrietvelde}}, \bibinfo {author} {\bibfnamefont {P.~A.}\ \bibnamefont
			{Hoehn}}, \bibinfo {author} {\bibfnamefont {F.}~\bibnamefont {Giacomini}}, \
		and\ \bibinfo {author} {\bibfnamefont {E.}~\bibnamefont {Castro-Ruiz}},\
	}\href@noop {} {\bibfield  {journal} {\bibinfo  {journal} {Quantum}\ }\textbf
		{\bibinfo {volume} {4}},\ \bibinfo {pages} {225} (\bibinfo {year}
		{2020})}\BibitemShut {NoStop}%
	\bibitem [{\citenamefont {Mersini-Houghton}(2014)}]{MersiniHorizonAvoidance}%
	\BibitemOpen
	\bibfield  {author} {\bibinfo {author} {\bibfnamefont {L.}~\bibnamefont
			{Mersini-Houghton}},\ }\href {\doibase
		https://doi.org/10.1016/j.physletb.2014.09.018} {\bibfield  {journal}
		{\bibinfo  {journal} {Phys. Lett. B}\ }\textbf {\bibinfo {volume} {738}},\
		\bibinfo {pages} {61 } (\bibinfo {year} {2014})}\BibitemShut {NoStop}%
\end{thebibliography}
%


\end{document}